\documentclass[longauth]{aa}

\usepackage[switch]{lineno}
\usepackage{graphicx}
\usepackage{arydshln}
\usepackage{subcaption}
\usepackage{hyperref}
\hypersetup{colorlinks=true,linkcolor=blue,filecolor=blue,urlcolor=blue,citecolor=blue}
\usepackage{appendix}
\usepackage{comment}
\usepackage{anyfontsize}
\usepackage{txfonts}
\makeatletter
\renewcommand*\aa@pageof{, page \thepage{} of \pageref*{LastPage}}
\makeatother

\begin{document}

   \title{First characterization of the emission behavior of Mrk~421 from radio to VHE gamma rays with simultaneous X-ray polarization measurements}
   \titlerunning{Radio to TeV observations of Mrk~421 during the \textit{IXPE} campaigns}
%
\author{\normalsize S.~Abe\inst{1} \and
J.~Abhir\inst{2} \and
V.~A.~Acciari\inst{3} \and
I.~Agudo\inst{4} \and
T.~Aniello\inst{5} \and
S.~Ansoldi\inst{6,44} \and
L.~A.~Antonelli\inst{5} \and
A.~Arbet Engels\inst{7}\thanks{Corresponding authors. E-mail: \href{mailto:contact.magic@mpp.mpg.de}{contact.magic@mpp.mpg.de}}\and
C.~Arcaro\inst{8} \and
M.~Artero\inst{9}
% ...
}
\institute { Japanese MAGIC Group: Institute for Cosmic Ray Research (ICRR), The University of Tokyo, Kashiwa, 277-8582 Chiba, Japan
\and ETH Z\"urich, CH-8093 Z\"urich, Switzerland
\and Instituto de Astrof\'isica de Canarias and Dpto. de  Astrof\'isica, Universidad de La Laguna, E-38200, La Laguna, Tenerife, Spain
\and Instituto de Astrof\'isica de Andaluc\'ia-CSIC, Glorieta de la Astronom\'ia s/n, 18008, Granada, Spain
\and National Institute for Astrophysics (INAF), I-00136 Rome, Italy
\and Universit\`a di Udine and INFN Trieste, I-33100 Udine, Italy
\and Max-Planck-Institut f\"ur Physik, D-80805 M\"unchen, Germany
\and Universit\`a di Padova and INFN, I-35131 Padova, Italy
\and Institut de F\'isica d'Altes Energies (IFAE), The Barcelona Institute of Science and Technology (BIST), E-08193 Bellaterra (Barcelona), Spain
\and Croatian MAGIC Group: University of Zagreb, Faculty of Electrical Engineering and Computing (FER), 10000 Zagreb, Croatia
\and IPARCOS Institute and EMFTEL Department, Universidad Complutense de Madrid, E-28040 Madrid, Spain
\and Centro Brasileiro de Pesquisas F\'isicas (CBPF), 22290-180 URCA, Rio de Janeiro (RJ), Brazil
\and University of Lodz, Faculty of Physics and Applied Informatics, Department of Astrophysics, 90-236 Lodz, Poland
\and Centro de Investigaciones Energ\'eticas, Medioambientales y Tecnol\'ogicas, E-28040 Madrid, Spain
\and Departament de F\'isica, and CERES-IEEC, Universitat Aut\`onoma de Barcelona, E-08193 Bellaterra, Spain
\and Universit\`a di Pisa and INFN Pisa, I-56126 Pisa, Italy
\and Universitat de Barcelona, ICCUB, IEEC-UB, E-08028 Barcelona, Spain
\and Department for Physics and Technology, University of Bergen, Norway
\and INFN MAGIC Group: INFN Sezione di Torino and Universit\`a degli Studi di Torino, I-10125 Torino, Italy
\and INFN MAGIC Group: INFN Sezione di Catania and Dipartimento di Fisica e Astronomia, University of Catania, I-95123 Catania, Italy
\and INFN MAGIC Group: INFN Sezione di Bari and Dipartimento Interateneo di Fisica dell'Universit\`a e del Politecnico di Bari, I-70125 Bari, Italy
\and Croatian MAGIC Group: University of Rijeka, Faculty of Physics, 51000 Rijeka, Croatia
\and Technische Universit\"at Dortmund, D-44221 Dortmund, Germany
\and University of Geneva, Chemin d'Ecogia 16, CH-1290 Versoix, Switzerland
\and Japanese MAGIC Group: Physics Program, Graduate School of Advanced Science and Engineering, Hiroshima University, 739-8526 Hiroshima, Japan
\and Deutsches Elektronen-Synchrotron (DESY), D-15738 Zeuthen, Germany
\and Armenian MAGIC Group: ICRANet-Armenia, 0019 Yerevan, Armenia
\and Croatian MAGIC Group: University of Split, Faculty of Electrical Engineering, Mechanical Engineering and Naval Architecture (FESB), 21000 Split, Croatia
\and Universit\"at W\"urzburg, D-97074 W\"urzburg, Germany
\and Croatian MAGIC Group: Josip Juraj Strossmayer University of Osijek, Department of Physics, 31000 Osijek, Croatia
\and Finnish MAGIC Group: Finnish Centre for Astronomy with ESO, University of Turku, FI-20014 Turku, Finland
\and Japanese MAGIC Group: Department of Physics, Tokai University, Hiratsuka, 259-1292 Kanagawa, Japan
\and Universit\`a di Siena and INFN Pisa, I-53100 Siena, Italy
\and Saha Institute of Nuclear Physics, A CI of Homi Bhabha National Institute, Kolkata 700064, West Bengal, India
\and Inst. for Nucl. Research and Nucl. Energy, Bulgarian Academy of Sciences, BG-1784 Sofia, Bulgaria
\and Japanese MAGIC Group: Department of Physics, Yamagata University, Yamagata 990-8560, Japan
\and Finnish MAGIC Group: Space Physics and Astronomy Research Unit, University of Oulu, FI-90014 Oulu, Finland
\vfill\null
\and Japanese MAGIC Group: Chiba University, ICEHAP, 263-8522 Chiba, Japan
\and Japanese MAGIC Group: Institute for Space-Earth Environmental Research and Kobayashi-Maskawa Institute for the Origin of Particles and the Universe, Nagoya University, 464-6801 Nagoya, Japan
\and Japanese MAGIC Group: Department of Physics, Kyoto University, 606-8502 Kyoto, Japan
\and INFN MAGIC Group: INFN Sezione di Perugia, I-06123 Perugia, Italy
\and INFN MAGIC Group: INFN Roma Tor Vergata, I-00133 Roma, Italy
\and Japanese MAGIC Group: Department of Physics, Konan University, Kobe, Hyogo 658-8501, Japan
\and also at International Center for Relativistic Astrophysics (ICRA), Rome, Italy
% ...
}

   \date{Received XX XX, 2023; accepted XX XX, 2023}

 
  \abstract
   {}
   { We perform the first broadband study of Mrk~421 from radio to TeV gamma rays with simultaneous measurements of the X-ray polarization from \textit{IXPE}. }
   {The data were collected within an extensive multiwavelength campaign organized between May and June 2022 using MAGIC, \textit{Fermi}-LAT, \textit{NuSTAR}, \textit{XMM-Newton}, \textit{Swift}, and several optical and radio telescopes to complement \textit{IXPE}. }
   {During the \textit{IXPE} exposures, the measured 0.2–1 TeV flux is close to the quiescent state and ranges from 25\% to 50\% of the Crab Nebula without intra-night variability. Throughout the campaign, the VHE and X-ray emission are positively correlated at a $4\sigma$ significance level. The IXPE measurements unveil a X-ray polarization degree that is a factor of 2-5 higher than in the optical/radio bands; that implies an energy-stratified jet in which the VHE photons are emitted co-spatially with the X-rays, in the vicinity of a shock front. The June 2022 observations exhibit a rotation of the X-ray polarization angle. Despite no simultaneous VHE coverage being available during a large fraction of the swing, the \textit{Swift}-XRT monitoring unveils an X-ray flux increase with a clear spectral hardening. It suggests that flares in high synchrotron peaked blazars can be accompanied by a polarization angle rotation, as observed in some flat spectrum radio quasars. Finally, during the polarization angle rotation, \textit{NuSTAR} data reveal two contiguous spectral hysteresis loops in opposite directions (clockwise and counter-clockwise), implying important changes in the particle acceleration efficiency on $\sim$hour timescales. }
   {}


   \maketitle
%


\section{Introduction} \label{sec:intro}
Blazars are a class of jetted active galactic nuclei (AGN) where the relativistic plasma jet is oriented at a small angle to the line of sight from Earth. They emit across the full electromagnetic spectrum from radio to very-high-energy gamma rays (VHE; E$\,>\,$100$\,$GeV). Blazars with no or very faint emission lines in the optical band are referred to as BL Lac type objects~\citep{1995PASP..107..803U}.\par

The spectral energy distribution (SED) of BL Lac type objects is dominated by the non-thermal radiation emission from the jet. The SED shows two large components, one peaking from radio to X-rays and a second component located in the gamma rays. It is widely accepted, based on spectral and polarization characteristics, that the first component originates from synchrotron radiation produced by relativistic electrons/positrons in the magnetic field within the jet. The exact origin of the second component is ambiguous to determine and still under debate. A possible scenario is electron inverse Compton (IC) scattering on synchrotron photons making up the first component, labelled as synchrotron self-Compton (SSC) model \citep{1992ApJ...397L...5M, 1999ApJ...521..145M}. In some cases, an additional target photon field for IC scattering is introduced to properly describe the SED of BL Lacs \citep[e.g.][]{1999ApJ...521..145M, 2013ApJ...768...54B}. Scenarios involving hadronic particles also provide possible explanations for the gamma-ray emission \citep{1993A&A...269...67M, 2015MNRAS.448..910C}. A common approach for classifying BL Lac type objects is by the peak frequency of their synchrotron component \citep{1995PASP..107..803U,2017A&ARv..25....2P}. Following the nomenclature of \citet{2010ApJ...716...30A}, blazars showing a synchrotron peak frequency $\nu_s<10^{14}\,$Hz are labelled low synchrotron peaked blazars (LSPs). Intermediate synchrotron peaked blazars (ISPs) show peak frequencies of $10^{14}\,\textrm{Hz}<\nu_s<10^{15}\,$Hz. Blazars with $\nu_s>10^{15}\,\textrm{Hz}$ are defined as high synchrotron peaked blazars (HSPs).\par

Markarian 421 (Mrk$\,$421; RA=11$^\text{h}$4$^{\prime}$27.31$^{\prime\prime}$, Dec=38$^{\circ}$12$^{\prime}$31.8$^{\prime\prime}$, J2000, z=0.031) is an archetypal HSP and among the closest and most extensively studied extragalactic sources in the VHE sky \citep[e.g.,][]{2009ApJ...695..596H,2016ApJ...819..156B,2021MNRAS.504.1427A}. Nevertheless, the exact processes for the acceleration of high-energy particles and the resulting emission mechanisms in Mrk~421 and blazars generally remain unclear. One promising approach to test acceleration and emission scenarios in HSPs is to measure the linear polarization throughout the spectrum~\citep{1985ApJ...298..114M, 2013ApJ...774...18Z, 2021Galax...9...37T}. Polarization measurements also provide important clues about the magnetic field ordering.\par

Prior blazar polarization measurements fell short of HSP synchrotron peak frequencies, extending only up to optical frequencies. Optical polarization measurements are thus not sufficient to probe the most energetic electrons freshly accelerated inside the jet. Since December 9, 2021, the Imaging X-ray Polarimetry Explorer (\textit{IXPE}) has been in orbit~\citep{2022HEAD...1930101W} and is able to perform measurements of the linear polarization of blazars between 2-8$\,$keV. The first detection of X-ray polarization from the blazar Markarian~501 (Mrk~501) by \textit{IXPE} was reported in~\citet{IXPE_Mrk501}. A high degree of linear polarization at the level of 10\% was detected without significant polarization variability. The X-ray polarization was in fact found to be significantly higher that in the optical and radio bands. These properties suggest a shock acceleration model with an energy-stratified electron population. In May and June 2022, \textit{IXPE} observed Mrk~421. Part of the results were published in \citet{IXPE_Mrk421} and \citet{mrk421_ixpe_rotation}. Similarly to Mrk~501, a high degree of linear polarization was detected in the X-ray compared to the optical and radio.\par

Starting in the year 2009, the blazar Mrk\,421 has been the focus of a multi-year program consisting in half-year dedicated observations with a number of instruments covering the broadband emission from radio to VHE gamma rays, with the first publication of this extensive observation program being \citet{abdo:2011}. Triggered by the planned observations of Mrk\,421 with \textit{IXPE}, the multi-instrument observations related to the extensive campaign on Mrk\,421 were intensified during (as well as before and after) the times when IXPE observed Mrk\,421. This intensified monitoring was particularly important for the Florian Goebel Major Atmospheric Gamma Imaging Cherenkov (MAGIC). In this work we present the first observations of a blazar in VHE gamma-rays accompanied by simultaneous X-ray polarization measurements. We have coordinated observations by a great number of instruments further complementing the \textit{IXPE} and VHE measurements with detailed coverage in X-rays by the \textit{Neil Gehrels Swift Observatory (Swift)}, the \textit{X-ray Multi-Mirror Mission (XMM-Newton)} and the \textit{Nuclear Spectroscopic Telescope Array (NuSTAR)}. High-energy gamma-ray observations are provided by the Large Area Telescope (LAT) on board the \textit{Fermi Gamma-ray Space Telescope} (\textit{Fermi}-LAT).\par

This paper is structured as follows: In Sect.~\ref{sec:data_analysis} we describe the observations and data analysis conducted with the different instruments. In Sect.~\ref{sec:characterization_MWL} we provide a detailed characterization of the multiwavelength (MWL) emission during the \textit{IXPE} observations, focusing on the spectral evolution, polarization features and intra-night variability. In Sect.~\ref{sec:long_term_MWL} we investigate the MWL behavior and correlations across the full campaign spanning from May to June 2022. At last, in Sect.~\ref{sec:discussion} we summarize and discuss the experimental findings of this study.  

\section{Observations and data processing} \label{sec:data_analysis}

\subsection{MAGIC}
The MAGIC telescopes consist of two 17-meter IACTs (MAGIC I and MAGIC II) located at the Observatorio del Roque de los Muchachos (ORM, 28.762$^\circ$N 17.890$^\circ$W, 2200 m above sea level) on the Canary Island of La Palma. Since 2009, stereoscopic observations are performed enabling the detection of gamma rays with energies from about 30$\,$GeV up to $\gtrsim$100$\,$TeV~\citep{aleksic:2016, MAGICCrab100TeV}.\par 

During the full time period covered by this work, we observed Mrk\,421 for 20.2$\,$h in total. The analysis is performed using the MAGIC Analysis and Reconstruction Software, MARS~\citep{MARS,aleksic:2016}, in the zenith angle range between 5$^\circ$ and 62$^\circ$. After applying quality cuts to remove data taken at too high of a zenith angle and during adverse weather conditions, 17.3$\,$h of data remained. The data were taken under low moonlight conditions, thus limiting contamination from night sky background light \citep{2017APh....94...29A}. Thanks to the brightness and proximity of Mrk\,421, two separate light curves can be obtained in the VHE regime covering an energy range from 0.2-1$\,$TeV and above 1$\,$TeV. The former light curve only contains data taken with a zenith angle of up to 50$^\circ$ due to the increasing energy threshold at larger zeniths~\citep{aleksic:2016}, while the latter includes the entire zenith range. \par 
The spectral analysis of the MAGIC data is performed by fitting the data with a log-parabolic model defined as follows: 
\begin{equation} \label{eq:log_par}
    \frac{\mathrm{d} N}{\mathrm{d} E} = f_0 \left( \frac{E}{E_0} \right)^{\alpha - \beta \, \text{log}_{10} \left( E/E_0 \right) }
\end{equation}
The normalization constant is given by $f_0$, $\alpha$ is the photon index at a normalization energy $E_0$, and $\beta$ is the curvature parameter. For the normalization energy, $E_0$, a fixed value of 300$\,$GeV is chosen. Flux points are obtained by performing the Tikhonov unfolding procedure as described in~\citet{unfolding}. All obtained parameters and flux points are corrected for the extragalactic background light (EBL) absorption following the model of~\citet{dominguez2011}.

\par

\subsection{\textit{Fermi}-LAT}

The LAT instrument is a pair-conversion telescope on board the \textit{Fermi} satellite \citep{2009ApJ...697.1071A,2012ApJS..203....4A} surveying the gamma-ray sky in the 20\,MeV to $>300$\,GeV energy range. For this work, we perform an unbinned-likelihood analysis using tools from the \texttt{FERMITOOLS} software\footnote{\url{https://fermi.gsfc.nasa.gov/ssc/data/analysis/}} v2.0.8. We use the instrument response function \texttt{P8R3\_SOURCE\_V2} and the diffuse background models\footnote{\url{http://fermi.gsfc.nasa.gov/ssc/data/access/lat/\\BackgroundModels.html}} \texttt{gll\_iem\_v07} and \texttt{iso\_P8R3\_SOURCE\_V3\_v1}.\par 

We select \texttt{Source} class events between 0.3\,GeV and 300\,GeV in a circular region of interest (ROI) with a radius of $20^\circ$ around Mrk\,421. The events with a zenith angle $>90^\circ$ are discarded to limit the contribution from limb gamma rays. To build the source model, we include all sources from the fourth Fermi-LAT source catalogue Data Release 2 \citep[4FGL-DR2;][]{2020ApJS..247...33A, 2020arXiv200511208B} that are found within the ROI plus an annulus of $5^\circ$. Mrk\,421 is modelled with a simple power-law function. In order to build light curves, the source model is fitted to the data by letting free to vary the normalization and the spectral parameters of all sources within $7^\circ$ of the target. Above $7^\circ$, all spectral parameters are fixed to the 4FGL-DR2 values. The normalizations of the diffuse background components are left as free parameters. When the fit does not converge, the model parameters are fixed to the 4FGL-DR2 values for sources detected with a test statistic \citep[TS;][]{1996ApJ...461..396M} below 4. If after that the fit still does not converge, we gradually increase the TS threshold below which the model parameters are fixed, until convergence is achieved.\par  

We produced a light curve in the 0.3-300\,GeV\footnote{The threshold energy of 0.3\,GeV was preferred over the usual 0.1\,GeV in order to exploit the improved angular resolution of \textit{Fermi}-LAT at higher energies. A higher energy threshold also reduces background contamination, which leads to an overall improvement of the signal-to-noise ratio for hard sources such has Mrk~421 (photon index~$>-2$).} band using 3-day time bins. In all time bins, the source is detected with TS~$>25$ (i.e., $>5\sigma$). Finally, we computed a SED around each \textit{IXPE} observation by averaging the data over 7 days. This time bin choice is a good compromise solution, given the flux variability observed in the light curves, and the limited sensitivity of LAT to measure gamma-ray spectra over short time intervals.\par 

\subsection{\textit{NuSTAR}}

This work comprises two multi-hour exposures from the  Nuclear Spectroscopic Telescope Array \citep[\textit{NuSTAR}][]{2013ApJ...770..103H}, which consists of two co-aligned X-ray telescopes focusing on two independent focal plane modules, FPMA and FPMB. The instrument provides unprecedented sensitivity in the 3-79\,keV band. The observations took place on 4\textsuperscript{th}-- 5\textsuperscript{th} June 2022 (MJD~59734 to MJD~59735) and 7\textsuperscript{th}-- 8\textsuperscript{th} June 2022 (MJD~59737 to MJD~59738; observation ID 60702061002 and 60702061004, respectively), with a total exposure time of 21\,ks and 23\,ks, respectively. The raw data are processed using the \textit{NuSTAR} Data Analysis Software (NuSTARDAS) package v.2.1.1 and CALDB version 20220912. The events are screened in the \texttt{nupipeline} process with the flags \texttt{tentacle=yes} and \texttt{saamode=optimized} in order to remove any potential background increase caused by the South Atlantic Anomaly passages. The source counts are obtained from a circular region centered around Mrk~421 with a radius of $\approx140''$. The background events are extracted from a source-free nearby circular region having the same radius. The spectra are then grouped with the \texttt{grppha} task to obtain at least 20 counts in each energy bin. \par 

For both exposures, the source spectra dominate over the background up to roughly $\approx 30$\,keV. Hence, in this work we decide to quote fluxes only up to 30\,keV, and in two separate energy bands: 3-7\,keV and 7-30\,keV. The best-fit spectral parameters averaged over the respective observations are obtained in the full \textit{NuSTAR} band-pass, 3-79\,keV. We fit the spectra using \texttt{XSPEC} \citep{1996ASPC..101...17A} assuming a log-parabolic function with a normalization energy fixed to 1\,keV. A simple power-law model provides a significantly worse description of the spectra (at a level $>5\sigma$ based on the $\chi^2$) and a curvature is detected during both observations. Here, and for the rest of the X-ray analysis performed in this work, a photoelectric absorption component is added to the model assuming an equivalent hydrogen column density fixed to $N_{\rm H}=1.34\times 10^{20}$\,cm$^{-2}$ \citep{2016A&A...594A.116H}. The fluxes and spectral parameters are computed by fitting simultaneously FPMA and FPMB. The cross-calibration factor between the two focal plane modules is for all bins within 0.95 and 1.05, thus well inside the expected systematics \citep{2015ApJS..220....8M}.

\subsection{\textit{Swift}-XRT}

We organized several X-ray pointings from the \textit{Swift} X-ray Telescope \citep[XRT;][]{2005SSRv..120..165B}. A special effort was put to schedule the observations simultaneously to the MAGIC exposures. The \textit{Swift}-XRT observations were performed in both Windowed Timing (WT) and Photon Counting (PC) readout modes. We processed the data using the XRTDAS software package (v.3.7.0) developed by the ASI Space Science Data Center\footnote{\url{https://www.ssdc.asi.it/}} (SSDC), released by the NASA High Energy Astrophysics Archive Research Center (HEASARC) in the HEASoft package (v.6.30.1). In order to calibrate and clean the events, data were reprocessed with the \texttt{xrtpipeline} script and using calibration files from \textit{Swift}-XRT CALDB (version 20210915) within the \texttt{xrtpipeline}.\par

For each observation, the X-ray spectrum was extracted from the calibrated and cleaned event file. In both WT and PC modes, the events were selected within a circle of 20-pixel ($\sim$47 arcsec) radius. The background was then extracted from a nearby, source-free, circular region with a 40-pixel radius. The ancillary response files were generated with the \texttt{xrtmkarf} task applying corrections for PSF losses and CCD defects using the cumulative exposure map.\par 

The $0.3-10$\,keV source spectra were binned using the \texttt{grppha} task by requiring at least 20 counts per energy bin. We then used \texttt{XSPEC} using both a power-law and log-parabola models (with a pivot energy fixed at 1 keV). In the vast majority of the observations, the statistical preference for a log-parabola model is significant ($>5\sigma$). The fluxes were extracted in the 0.3-2\,keV, and 2-10\,keV energy bands.

\subsection{\textit{XMM-Newton}}

The {\it XMM-Newton} observatory carries on board several coaligned X-ray instruments: the European Photon Imaging Camera (EPIC) and two Reflection Grating Spectrometers \citep[RGS1 and RGS2,][]{2001A&A...365L...1J}. The EPIC cameras consist of two Metal Oxide Semiconductors \citep[EPIC-MOS1 \& MOS2,][]{2001A&A...365L..27T} and one pn junction \citep[EPIC-pn,][]{2001A&A...365L..18S} CCD arrays operating in the 0.2--10 keV energy band. All {\it XMM-Newton} observations presented in this paper were taken with the EPIC camera under TIMING mode with the THICK filter. Data are available in the EPIC-pn and EPIC-MOS2 cameras. Observing times per observation range between $\sim$17~ksec and $\sim$47~ksecs. Our sample was analyzed using the {\it XMM-Newton} standard Science Analysis System \citep[SAS, v20.0.0;][]{2004ASPC..314..759G} and most updated calibration files. Event lists are produced for the two EPIC cameras following the standard SAS reduction procedure. Periods of high-background activity are removed following the standard method \citep{2002A&A...389...93L}.\par 

The source and background regions for the EPIC-pn and EPIC-MOS2 cameras are extracted following the same method as described in \citet{2021A&A...655A..48D}. We extract spectra in the full energy range (0.2--10~keV) with an energy resolution of 5~eV. The spectra are re-binned in order not to over sample the intrinsic energy resolution of the EPIC cameras by a factor larger than 3, while making sure that each spectral channel contains at least 25 background-subtracted counts. Spectral fits are performed with the \texttt{XSPEC} package \citep{1996ASPC..101...17A} in the energy range 0.6--10~keV (a minimum fit energy of $\approx0.6$\,keV is recommended by the official SAS documentation\footnote{\url{https://www.cosmos.esa.int/web/xmm-newton/calibration-documentation}} for TIMING mode observations to avoid low energy noise distorting the spectra). For every observation, we perform spectral fits and derive spectral parameters from the combined EPIC instruments available (i.e. EPIC-pn and EPIC-MOS2). All spectra are fitted using a log-parabola model (with a pivot energy set at 1\,keV).\par 

The most updated comparison of X-ray satellite observations shows that EPIC-pn data slightly differ from the \textit{NuSTAR} and \textit{Swift} data both in flux and slope \citep{2017AJ....153....2M}. The EPIC-pn fluxes are 
significantly lower than the \textit{NuSTAR} fluxes, typically by the order of 20\%. Although, based on the analysis performed so far, it is not possible to elucidate which instrument recovers the correct X-ray fluxes, the XMM-Newton Science Operation Center has proposed a correction to the {\it XMM-Newton} EPIC data that can be applied for observations performed simultaneously with \textit{NuSTAR}. This correction has been applied to all the EPIC data that has simultaneous data with {\it NuSTAR}. The correction is applied in the ARF generation and can be done using the standard SAS task {\tt arfgen} including the parameter {\tt applyabsfluxcorr=yes}\footnote{\url{https://xmmweb.esac.esa.int/docs/documents/CAL-TN-0230-1-3.pdf}}.

\subsection{\textit{IXPE}}
The \textit{IXPE} telescope \citep{2022HEAD...1930101W} is the first instrument capable of resolving the X-ray polarization degree and angle in blazars. Here, we exploit the first three \textit{IXPE} observations of Mrk~421, which took place in the first half of 2022 and were accompanied by the simultaneous MAGIC monitoring. The first observation spanned from May 4\textsuperscript{th} 2022 10:00 UTC (MJD~59703) until May 6\textsuperscript{th} 2022 11:10 UTC (MJD~59705), for a total exposure of 97\,ks. The two additional observations took place in June 2022: from June 4\textsuperscript{th} 2022 10:56 UTC until June 6\textsuperscript{th} 2022 11:08 UTC (MJD~59734 to MJD~59736; 96\,ksec exposure time), and from June 7\textsuperscript{th} 2022 08:49 UTC until June 9\textsuperscript{th} 2022 09:51 UTC (MJD~59737 to MJD~59739; 86\,ksec exposure time). All results shown in this paper were taken from \citet{IXPE_Mrk421} (May observation) and \citet{mrk421_ixpe_rotation} (June observations). We refer the reader to the latter works for details about the analysis procedure.\par

During the first \textit{IXPE} observation, in May 2022, no variability of the polarization degree and angle is measured \citep{IXPE_Mrk421}, and the values averaged over the full exposure are considered. Regarding the two observations in June 2022, the polarization angle exhibits a large rotation \citep{mrk421_ixpe_rotation} at a speed of $80\pm9 \, ^{\circ}$/day (4\textsuperscript{th}--6\textsuperscript{th} June 2022; MJD~59734 to MJD~59736) and $91\pm8 \, ^{\circ}$/day (7\textsuperscript{th}--9\textsuperscript{th} June 2022; MJD~59737 to MJD~59739). The rotation is evident when considering the data binned in 3\,hour intervals. Based on simulations, \citet{mrk421_ixpe_rotation} estimated that the probability to detect these rotations due to random walks is about 2\%, and thus, it is highly unlikely to have occurred by chance. As described in \citet{mrk421_ixpe_rotation}, the polarization degree remains consistent with a constant behavior hypothesis.

\subsection{\textit{Swift}-UVOT}

We obtained a coverage in the UV band from the \textit{Swift} UV/Optical Telescope \citep[UVOT,][]{2005SSRv..120...95R}. We consider observations between April 26\textsuperscript{th} 2022 (MJD~59695) and June 27\textsuperscript{th} 2022 (MJD~59757) with the W1, M2 and W2 filters. We selected a sample of 43 observations of Mrk 421 from the official data archive, by applying standard quality checks to all observations in the chosen time interval, excluding those with unstable attitude or affected by contamination from a nearby star light (51 UMa). For each observation, we performed photometry over the total exposures in each filter. The same apertures for source counts (the standard with 5\,arcsec radius) and background estimation (mostly three-four circles of $\sim$16\,arcsec radii off the source) were applied to all. We used the official software included in the HEAsoft 6.23 package, from HEASARC, to perform the photometry extraction and then applied the official calibrations \citep{2011AIPC.1358..373B} included in the recent CALDB release (20201026). Finally, we dereddened source fluxes according to a mean interstellar extinction curve \citep{1999PASP..111...63F} and the mean Galactic $E(B - V)$ value of 0.0123 mag \citep{1998ApJ...500..525S, 2011ApJ...737..103S}.

\subsection{Optical observations}

In the optical, we exploit R-band photometric and polarimetric observations from the RoboPol \citep[Skinakas observatory, Greece;][]{2014MNRAS.442.1706K, 2019MNRAS.485.2355R}, Nordic Optical Telescope (NOT; ORM, Spain) and KANATA (Higashi-Hiroshima observatory, Japan) telescopes. We also make use of H-band (infrared; IR) data from the Perkins telescope (Perkins Telescope observatory, Flagstaff, AZ). All the latter data were published in \citet{IXPE_Mrk421} and \citet{mrk421_ixpe_rotation}, where more details on the analysis procedures can be found. Additional polarimetric and photometric observations of the source in the Johnson Cousins R-band band were performed at Sierra Nevada Observatory, Granada, Spain, with a four-unit polarized filter-wheel mounted at the 0.9 m\,telescope (here after dubbed T90). Unpolarized standard stars were also observed to compute the instrumental polarization that was subtracted from the actual data. Standard pre-reduction and analysis steps were performed.\par

All the polarization and photometric data were corrected for the contribution of the host galaxy using the host fluxes reported in \citet{2007A&A...475..199N}. The intrinsic polarization degree was obtained using the following formula: $P_{deg,intr} = P_{deg, obs} \cdot I / (I - I_{host})$, where $P_{deg, obs}$ the observed polarization degree, $I$ the observed flux and $I_{host}$ the host flux. Finally, the flux densities were also corrected for a galactic extinction of $0.033$\,mag according to the NASA/IPAC Extragalactic Database (NED)\footnote{\url{https://ned.ipac.caltech.edu/}}.

\subsection{Radio observations}

We collected data in the microwave band at 3.5\,mm (86.24\,GHz) and 1.3\,mm (230\,GHz) wavelengths with the 30\,m telescope of the Institut de Radioastronomie Millimetrique (IRAM) that is located at the Pico Veleta Observatory (Sierra Nevada, Granada, Spain). The observations were performed within the Polarimetric Monitoring of AGN at Millimeter Wavelengths (POLAMI) program\footnote{\url{https://polami.iaa.es/}} \citep{2018MNRAS.473.1850A, 2018MNRAS.474.1427A}. The four Stokes parameters (I, Q, U and V) were recorded simultaneously using the XPOL procedure \citep{2008PASP..120..777T}. The data reduction and calibration was achieved following the POLAMI procedure described in \citet{2018MNRAS.474.1427A}.\par 

Additional radio observations were performed by the Mets\"{a}hovi telescope. A detailed description of the data reduction and analysis can be found in \cite{Teraesranta1998}. In short, observations at 37~GHz are conducted using the 13.7~m Mets\"{a}hovi telescope. Under optimal conditions the detection limit of the telescope at 37~GHz is approximately 0.2 Jy. For the flux density, DR~21 is used as the primary calibrator, and NGC~7027, 3C~274 and 3C~84 are used as secondary calibrators. The flux density errors include the uncertainty in the absolute flux calibration as well as the root mean square of the measurement. We consider as detections only the observations with a signal-to-noise ratio greater than four.\par

Finally, we collected millimeter radio polarimetric measurements at 1.3~mm (approximately 230~GHz) with the Submillimeter Array \citep[SMA;][]{Ho2004}. The observations were conducted within the SMA Monitoring of AGNs with POLarization (SMAPOL) program in full polarization mode using SMA polarimeter \citep{Marrone2008} and SWARM correlator \citep{Primiani2016}. The polarized intensity, position angle, and polarization percentage were derived from the Stokes I, Q, and U visibilities and calibrated with the MIR software package\footnote{\url{https://lweb.cfa.harvard.edu/~cqi/mircook.html}} using MWC\,349\,A, Callisto (total flux calibrators) and 3C\,286 (polarized calibrator).

\begin{figure*}
  \centering
  \includegraphics[width=2\columnwidth]{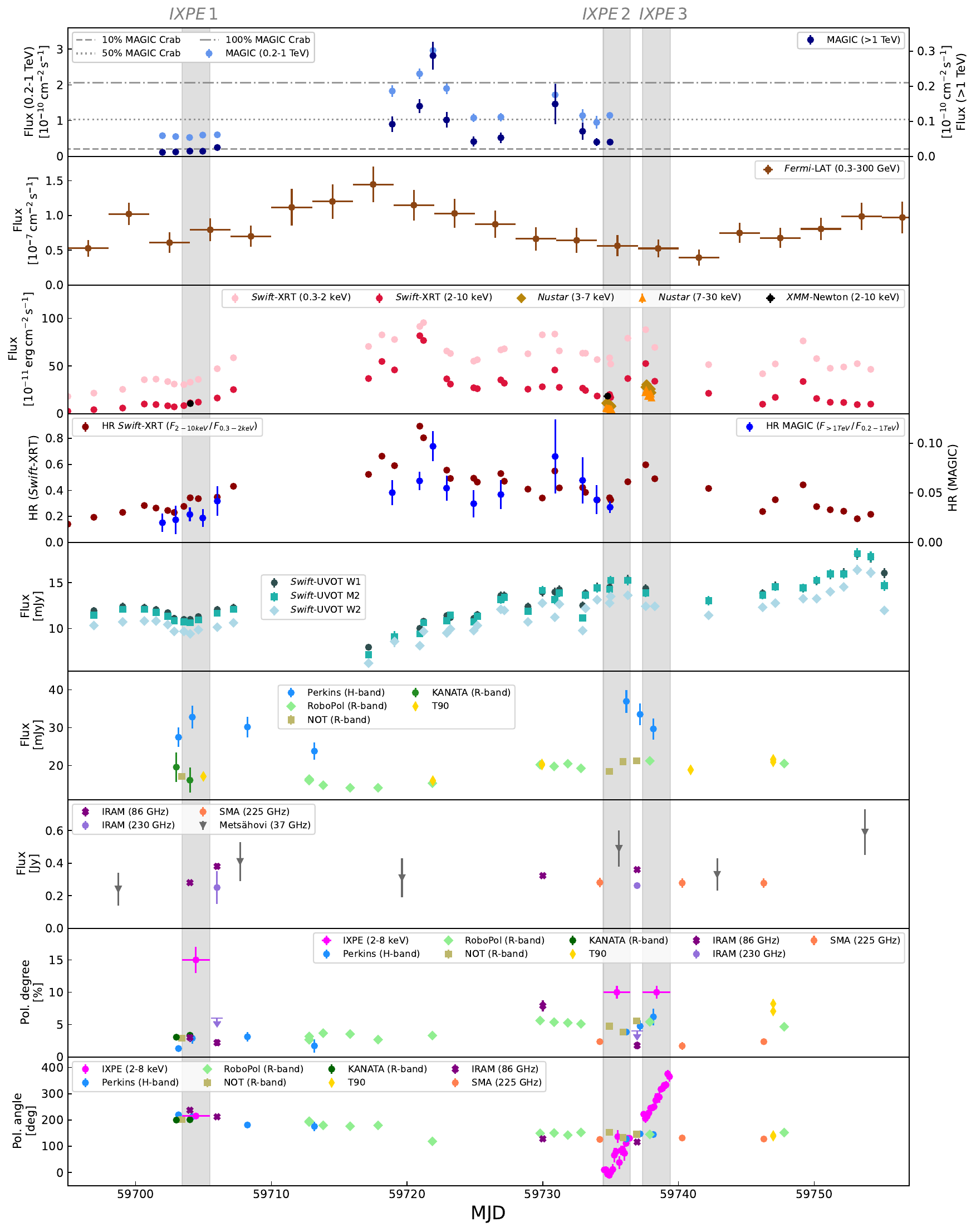}
 \caption{\small MWL light curve for Mrk\,421 covering the whole campaign from MJD~59695 (26\textsuperscript{th} April 2022) to MJD~59757 (27\textsuperscript{th} June 2022). The gray bands correspond to the three \textit{IXPE} observations. Top to bottom: MAGIC fluxes in daily bins for two energy bands (note the two different y-axes); \textit{Fermi}-LAT fluxes in 3\,day bins; X-ray fluxes in daily bins including \textit{Swift}-XRT, \textit{NuSTAR} and \textit{IXPE}; hardness ratio between the high- and low-energy fluxes of \textit{Swift}-XRT and between the two VHE bands of MAGIC (note the two different y-axes); optical R-band data from NOT, RoboPol KANATA; IR H-band data from Perkins; radio data from IRAM and SMA; polarization degree and polarization angle observations in the optical to radio from NOT, RoboPol, KANATA, Perkins, IRAM, SMA and in X-rays from \textit{IXPE}.}
 \label{fig:longterm_LC}
\end{figure*}

\section{Characterization of the VHE to radio behavior during \textit{IXPE} observations} \label{sec:characterization_MWL}

Fig.~\ref{fig:longterm_LC} shows the MWL light curves from MJD~59695 (26\textsuperscript{th} April 2022) to MJD~59760 (30\textsuperscript{th} June 2022), which encompasses all \textit{IXPE} observing periods. In the top row, the VHE energy bands (0.2-1\,TeV and $>1$\,TeV) are shown. As previously mentioned, data observed at a zenith above 50$^{\circ}$ were excluded from the 0.2-1\,TeV energy band, while for the $>1$\,TeV fluxes no cut on the zenith distance was applied. The cut on the zenith distance is necessary because the energy threshold increases to above 0.2\,TeV for zenith angles greater than 50$^{\circ}$ , and hence we would introduce artificial downward fluctuations in the reported fluxes (e.g. by producing a light curve above 0.2\,TeV when using data with an energy threshold well above this energy). In any case, this selection cut only removes a small fraction of the data from the 0.2-1\,TeV light curve (it affects only three nights, removing a total of $\approx2$\,hrs), and no intra-night variability was found in any of the two bands. Thus, the slightly different underlying data selection does not affect in any significant manner the hardness ratio. Measurements from \textit{Fermi}-LAT in the 0.3-300\,GeV band are portrayed in the second panel from the top. The \textit{Fermi}-LAT fluxes are computed in 3-day bins, providing a good trade-off between flux uncertainty and temporal resolution. In X-rays, third panel, a dense temporal coverage is given by \textit{Swift}-XRT in two energy bands (0.3-2\,keV and 2-10\,keV). On selected days during the \textit{IXPE} observations, additional data by \textit{NuSTAR} and \textit{XMM-Newton} are available. We quantify the corresponding spectral evolution using the hardness ratio in X-rays, defined as the ratio of the $2-10$\,keV flux to the $0.3-2$\,keV flux, in the fourth panel. Additionally, the hardness ratio of the VHE data (defined as the ratio of the $>1$\,TeV flux to the $0.2-1$\,TeV flux) is shown. UV observations from \textit{Swift}-UVOT in the W1, M2 and W2 filters are shown in the fifth panel from the top. We complement the MWL light curves with further data in the optical/IR and radio, which are plotted in the sixth and seventh panel, respectively. The last two panels at the bottom of Fig.~\ref{fig:longterm_LC} display the evolution of the polarization degree and polarization angle in the radio, optical/IR and X-ray. \par 

\subsection{\textit{IXPE} observation in May 2022}\label{sec:IXPE_1_mwl}

The first observation of Mrk\,421 by \textit{IXPE} occurred between the 4\textsuperscript{th} and 6\textsuperscript{th} of May 2022 (MJD 59703.42 - MJD 59705.47) and is shown as the first grey band in Fig.~\ref{fig:longterm_LC}. Here and in the following, this epoch will be referred to as \textit{IXPE 1}.\par

The MAGIC telescopes achieved a continuous daily coverage over the entire \textit{IXPE} exposure. In both VHE energy bands, the flux exhibits a constant behavior throughout the specified time period, showing a flux slightly below 10\% of the emission of the Crab Nebula\footnote{The flux of the Crab Nebula used in this work is taken from ~\citet{aleksic:2016}} in the range above 1$\,$TeV, and around 25\% for the 0.2-1$\,$TeV range. We do not find significant flux or spectral variability on daily and sub-daily timescale. A simultaneous X-ray characterization is obtained thanks to \textit{Swift}-XRT as well as a long exposure from \textit{XMM-Newton} on MJD~59704 (May 5\textsuperscript{th} 2022). The flux in both energy bands of the \textit{Swift}-XRT instrument exhibits moderate daily variability. In the 0.3-2\,keV band, a flux increase at the level of 20\% is observed, while it is 40\% in the 2-10\,keV band. The hardness ratio rises from $0.23\pm0.01$ up to almost $0.35\pm0.01$, indicating a harder-when-brighter trend in agreement with previous observations of Mrk~421  \citep[see for instance][]{2015A&A...576A.126A, 2021MNRAS.504.1427A, 2021A&A...655A..89M}. Regarding the multi-hour \textit{XMM-Newton} pointing, the average 2-10\,keV flux (pink marker in Fig.~\ref{fig:longterm_LC}) is consistent with \textit{Swift}-XRT results. During the observation, little variability is observed. A 500\,s binned \textit{XMM-Newton} light curve is shown in Fig.~\ref{fig:xmm_lc_ixpe1} of Appendix~\ref{sec:appendix_xmm_light_curve}. The concurrent optical/IR (R-band and H-band) and radio flux data in Fig.~\ref{fig:longterm_LC} around \textit{IXPE~1} show small variability although the limited temporal coverage prevents a detailed variability characterization.\par 

The degree of polarization from radio to optical shows slightly fluctuating values around 3\%. The results of the \textit{IXPE} observation (taken from~\citet{IXPE_Mrk421}) show a much higher constant degree of polarization of $15\pm2\,\%$ in the X-ray band. The polarization angle determined by \textit{IXPE} is $215\pm4\,^{\circ}$ (or $35\pm4\,^{\circ}$, if one considers the $180^\circ$ ambiguity in polarization angle measurements) and is in agreement with the angles measured in radio to optical, which range from around 200$\,^{\circ}$ up to 230$\,^{\circ}$ and also remains constant throughout the observation period. \par

\begin{figure*}
\centering
  \includegraphics[width=2\columnwidth]{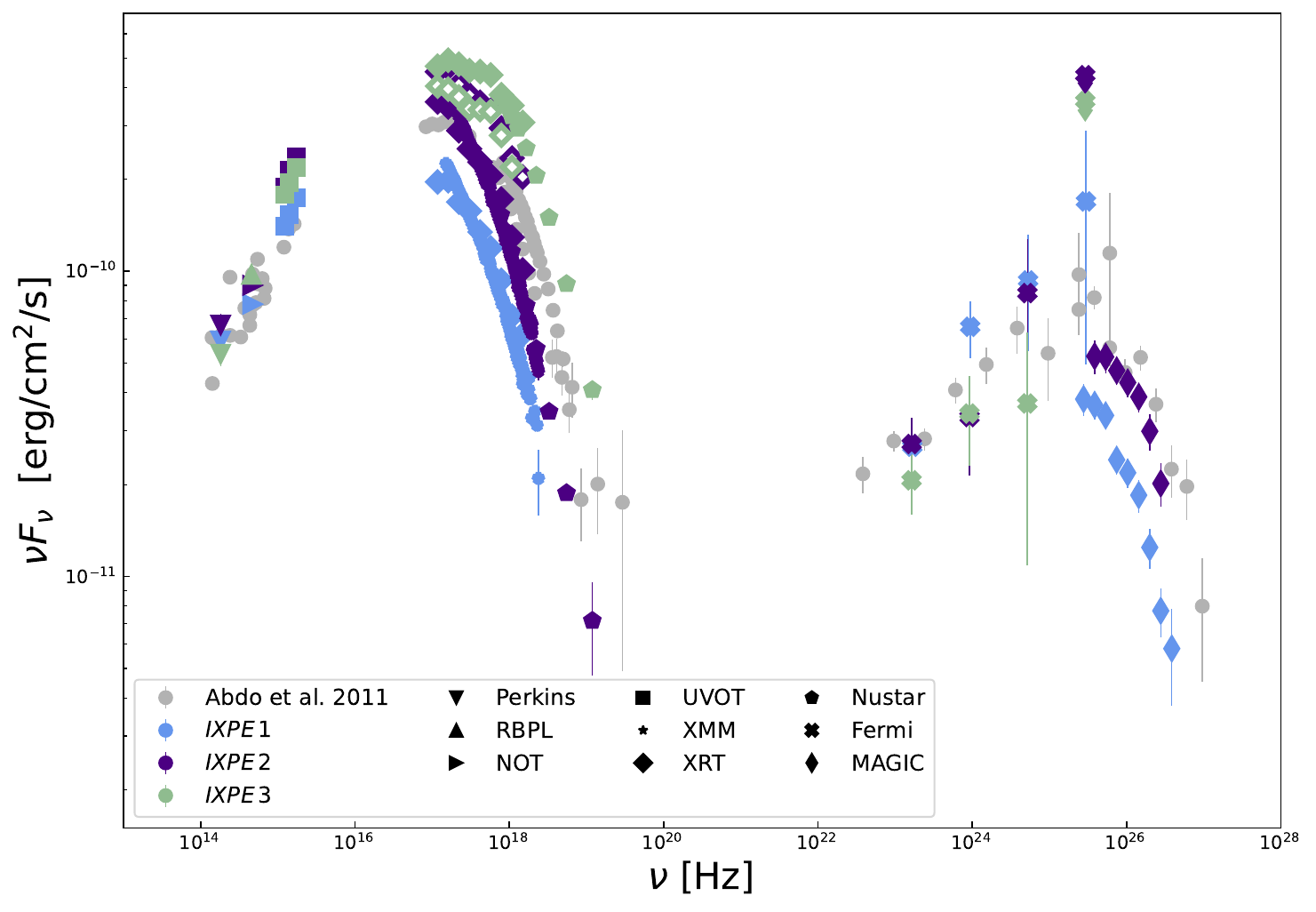}
  \caption{Broadband SED around the three \textit{IXPE} observations. Data from MAGIC were corrected for the EBL absorption using the model outlined in~\citet{dominguez2011}. In plain colored markers, the \textit{Swift}-XRT data correspond to the pointing that happened first within the \textit{IXPE} windows. The \textit{Swift}-XRT data in diamond markers with a facecolor in white are from the subsequent observation. For comparison, the SED of an average emission state of Mrk~421 from~\citet{abdo:2011} is shown in light grey.}
  \label{fig:SED}
\end{figure*}

\begin{center}
\begin{table*}[h]
\caption{Spectral parameters from the VHE and X-ray observations around the three \textit{IXPE} observing epochs.} 

\centering
\begin{tabular}{ c c | c c c }     
 & & \textit{IXPE~1} & \textit{IXPE~2} & \textit{IXPE~3} \\
\hline  \hline
MAGIC & MJD & 59703.5 to 59705.5 & 59734.5 to 59735.5 & --  \\
& Flux & $0.34\pm 0.01$ & $0.67 \pm 0.03$ & --\\
& $\alpha$ & $-2.64 \pm 0.06$ & $-2.30 \pm 0.08$ & -- \\
& $\chi^{2}$/d.o.f. & 25.0 / 13 & 2.5 / 9  & -- \\ \hline
\textit{Swift}-XRT & MJD & 59703.55$\quad$59704.02$\quad$59704.62 & 59734.92$\quad$59735.00$\quad$59736.27 & 59737.58$\quad$59738.25 \\
& Flux & $0.84^{+0.04}_{-0.03} \quad 1.13^{+0.04}_{-0.04} \quad 1.20^{+0.04}_{-0.04}$ & $2.01^{+0.05}_{-0.05}\quad 1.70^{+0.05}_{-0.05} \quad 3.68^{+0.08}_{-0.08}$ & $5.26^{+0.05}_{-0.06} \quad 3.40^{+0.12}_{-0.12}$ \\
& $\alpha$ & $-2.52^{+0.02}_{-0.02}$ \quad $-2.38^{+0.02}_{-0.02}$ \quad $-2.38^{+0.02}_{-0.02}$ & $-2.40^{+0.01}_{-0.01}$ \quad $-2.43^{+0.02}_{-0.02}$ \quad $-2.22^{+0.01}_{-0.01}$ &  $-2.07^{+0.01}_{-0.01}$ \quad $-2.19^{+0.02}_{-0.02}$  \\
& $\chi^{2}$/d.o.f. & 223.5/197$\quad$236.1/216$\quad$225.4/220 & 261.5/266$\quad$241.9/223$\quad$346.3/309 & 549.0/481$\quad$218.6/199  \\ \hline
\textit{XMM-Newton} & MJD & 59703.93 to 59704.13 & 59734.68 to 59735.11 & -- \\
& Flux & $1.056^{+0.002}_{-0.002}$ & $1.838^{+0.002}_{-0.002}$ & -- \\
& $\alpha$ & $-2.541^{+0.001}_{-0.001}$ & $-2.545^{+0.001}_{-0.001}$ & -- \\
& $\chi^{2}$/d.o.f. & 814.8/329  & 2082.13/345 & -- \\ \hline
\textit{NuSTAR} & MJD & -- & 59734.65 to 59735.11 & 59737.53 to 59738.04 \\
& Flux & -- & $0.968\pm0.004$ & $2.693\pm0.006$ \\
& $\alpha$ & -- & $-2.309\pm0.007$ & $-1.913\pm0.004$ \\
& $\chi^{2}$/d.o.f. & -- & 704.7/761 & 1143.8/1133 \\ 
\hline \hline
\end{tabular}
\tablefoot{The table contains four primary rows corresponding to the different instruments. The first subrow for an individual instrument shows the MJD of the observations performed during the three \textit{IXPE} observations given by the main columns. The second subrow contains the obtained fluxes in units of $10^{-10} \, \text{erg}\,\text{cm}^{-2} \text{s}^{-1}$ (for MAGIC the integrated photon flux between $200$\,GeV and $1$\,TeV is used, for \textit{Swift}-XRT and \textit{XMM-Newton} the flux between 2-10\,keV and for \textit{NuSTAR} the 3-7\,keV flux). The spectral index $\alpha$, assuming a log-parabola for MAGIC (with $\beta$ fixed to $0.50$ and reference energy of 300\,GeV) as well as for \textit{Swift}-XRT (with $\beta$ fixed to $0.29$ and reference energy of 1\,keV), \textit{XMM-Newton} (with $\beta$ fixed to $0.20$ and reference energy of 1\,keV), and \textit{NuSTAR} (with $\beta$ fixed to $0.45$ and reference energy of 1\,keV) is given in the third subrow. The last subrow gives the $\chi^{2}$/d.o.f.. Regarding \textit{XMM-Newton} and \textit{NuSTAR}, the parameters are obtained by fitting jointly the data from the available cameras onboard these observatories (i.e., EPIC-pn and EPIC-MOS2 for \textit{XMM-Newton}, FPMA and FPMB for \textit{NuSTAR}).}
\label{tab:IXPE1}
\end{table*}
\end{center}

\subsection{\textit{IXPE} observation in June 2022}\label{sec:IXPE_23_mwl}

The second and third \textit{IXPE} observations of Mrk\,421 were performed between the 4\textsuperscript{th} and 6\textsuperscript{th} of June 2022 (MJD 59734.46 - MJD 59736.46) and between the 7\textsuperscript{th} and 9\textsuperscript{th} of June 2022 (MJD 59737.36 - MJD 59739.41). In the following, the latter observing epochs are dubbed as \textit{IXPE~2} and \textit{IXPE~3}, respectively. These epochs are highlighted with vertical grey bands in Fig.~\ref{fig:longterm_LC}.\par 

MAGIC could only observe during the first day of the \textit{IXPE~2} period as well as  two days before, for a total of 3.3$\,$h of observation. Over the course of three days, the flux in the 0.2-1$\,$TeV band is close to $\approx$50\% of the Crab Nebula and $\approx$20\% above 1$\,$TeV, indicating about twice as much flux as during \textit{IXPE 1}.\par 

In X-rays, a significantly higher activity is also observed throughout the entire \textit{IXPE~2} and \textit{IXPE~3} windows with respect to \textit{IXPE~1}, and the source exhibits clear spectral and flux variability. Between the \textit{IXPE~2} epoch and the start of the \textit{IXPE~3} epoch, the 2-10\,keV flux shows a steady increase by a factor $\approx2.6$, together with a clear hardening of the emission that is highlighted by the hardness ratio evolution (a more detailed spectral analysis is presented in Sect.~\ref{sec:spectral_behavior}). The peak activity in the 2-10\,keV band is about five times the average flux level observed during \textit{IXPE~1}. Although this flux state is still below previous X-ray outbursts of Mrk~421 \citep[see for instance the March 2010 flare reported in ][]{2015A&A...578A..22A}, this activity is among the highest states recorded during 2022. The flux then decreases during the last \textit{Swift}-XRT observation simultaneous to \textit{IXPE~3}. The \textit{XMM-Newton} analysis confirms the higher X-ray activity compared to \textit{IXPE~1}. The observation took place at the beginning of the \textit{IXPE~2} epoch, slightly before the clear flux increase witnessed by \textit{Swift}-XRT. In addition to \textit{Swift}-XRT and \textit{XMM-Newton}, a precise hard X-ray characterization was obtained thanks to two multi-hour \textit{NuSTAR} exposures during both \textit{IXPE~2} and \textit{IXPE~3}. In the third panel from the top of Fig.~\ref{fig:longterm_LC}, we show the \textit{NuSTAR} fluxes in the 3-7\,keV and 7-30\,keV bands using 1\,hour time bins. For \textit{IXPE~2}, the \textit{NuSTAR} observation was simultaneous to MAGIC. The corresponding intra-night VHE versus X-ray correlation is investigated in Section~\ref{sec:intranight_ixpe2}. During both \textit{NuSTAR} pointings, a moderate flux change is observed on hour timescales (at the level of $30\%$). Nonetheless, a detailed study unveils spectral hysteresis patterns. This analysis is presented in Sect~\ref{sec:intranight_full_nustar}.\par 

Regarding the MeV-GeV band, the \textit{Fermi}-LAT analysis shows a similar flux state as during \textit{IXPE~1}, and is close to the average activity for Mrk~421 \citep{abdo:2011}. For the UV, optical, IR and radio emission, here also the emission does not reveal significant evolution with respect to \textit{IXPE~1}.\par 

The bottom panels of Fig.~\ref{fig:longterm_LC} show the evolution of the polarization degree and angle in the X-ray 2-8\,keV band \citep[pink markers; the data are taken from][]{mrk421_ixpe_rotation}. During \textit{IXPE~2} and \textit{IXPE~3}, the averaged degree is $10\pm1\,\%$. While the polarization degree is consistent with a constant behavior (see also Sect.~\ref{sec:intranight_full_nustar}), the polarization angle exhibits an evident rotation, which seems continuous between the two \textit{IXPE~2} and \textit{IXPE~3} epochs. During \textit{IXPE~2}, the angle rotates at an average angular velocity of $80\pm9 \, ^{\circ}$/day amounting to a total rotation of 120$\, ^{\circ}$. The rotation continued at a compatible rate of $91\pm8 \, ^{\circ}$/day during \textit{IXPE~3}, for a total rotation of 140$\, ^{\circ}$. The significant X-ray flux increase and spectral hardening measured by \textit{Swift}-XRT is thus accompanied by a rotation of the polarization angle. In Section~\ref{sec:intranight_full_nustar}, we investigate the short timescale spectral variability in the hard X-rays during the polarization angle rotation using simultaneous \textit{NuSTAR} data.\par 

It is interesting to note that at lower frequencies, in the radio/IR/optical bands, both the flux and polarization properties do not show any prominent variability. The polarization degree in the optical and IR fluctuates around 5\% while the radio polarization is slightly lower, around 2\% both for the 86\,GHz and 230\,GHz bands.\par

\subsection{Spectral evolution throughout the \textit{IXPE} observing epochs}
\label{sec:spectral_behavior}

Fig.~\ref{fig:SED} presents the simultaneous broadband SEDs during each of the \textit{IXPE} periods from the IR up to VHE gamma rays. In comparison, the average state of Mrk\,421 taken from \citet{abdo:2011} is plotted in light grey. Since the VHE flux level reported in~\citet{abdo:2011} is close to the average state found by Whipple over a time span of 14 years~\citep[45\% of the Crab Nebula flux,][]{2014APh....54....1A}, we consider the broadband SED of~\citet{abdo:2011} as an average activity state and use it as a reference for comparison. MAGIC VHE flux points from \textit{IXPE~1} are obtained by averaging all data within the corresponding \textit{IXPE} exposures since we find no significant spectral no flux variability. Regarding \textit{IXPE~2}, a single MAGIC observation is available and it took place at the beginning of the \textit{IXPE} window, while \textit{IXPE~3} is lacking VHE coverage (see Fig.~\ref{fig:longterm_LC} and previous section). The \textit{Fermi}-LAT SEDs are averaged over 7 days, centered around the \textit{IXPE} windows. In X-rays, for \textit{IXPE~1}, we show the \textit{Swift}-XRT SED on MJD~59704.02 (May 5\textsuperscript{th} 2022), which is close to the center of the \textit{IXPE} window and simultaneous to the \textit{XMM-Newton} observation. Regarding \textit{IXPE~2} and \textit{IXPE~3}, we plot for each epoch the \textit{Swift}-XRT SEDs that were first recorded within the \textit{IXPE} windows. The latter SEDs are also accompanied by simultaneous \textit{XMM-Newton} (for \textit{IXPE~2} only) and \textit{NuSTAR} data (for both \textit{IXPE~2} and \textit{IXPE~3}). We add \textit{Swift}-XRT SEDs corresponding to the last pointing before the end of the \textit{IXPE} windows in order to illustrate the daily timescale variability along the \textit{IXPE} exposure. For the optical and IR data, we use measurements that are the closest in time to each of the X-ray observations.\par 

Compared to the average state of~\cite{abdo:2011}, the \textit{IXPE~1} epoch (blue markers) displays a VHE and X-ray emission that is significantly lower. The X-ray SED is also softer, indicating a shift of the synchrotron peak towards lower frequencies. Based on a log-parabola fit in the SED space ($\nu F_{\nu} \propto 10^{-b (\log{(\nu/\nu_p)})^2}$), we derive a peak frequency located at $\nu_p = (2.00\pm0.07) \times 10^{16}\,$Hz, while the state from \cite{abdo:2011} suggests $\nu_p \approx 10^{17}$\,Hz. Throughout the \textit{IXPE~2} and \textit{IXPE~3}, Fig.~\ref{fig:SED} highlights clearly the spectral changes occurring during the polarization angle swing reported by \textit{IXPE}. At the beginning of \textit{IXPE~2} (plain violet color markers), the emission is roughly comparable to the typical state at all frequencies. Compared to \textit{IXPE~1}, the synchrotron peak frequency increases marginally to $\nu_p = (2.27\pm0.09) \times 10^{16}\,$Hz. The emission increases significantly during the subsequent X-ray SED, which shows a flux well above the typical state as well as an harder emission. The maximum observed brightness is reached at the start of \textit{IXPE~3} (green markers), which coincides with the second \textit{NuSTAR} observation and shows an enhanced emission state throughout the full synchrotron peak accompanied by a significant shift of the synchrotron peak towards a higher frequency ($\nu_p = (7.6\pm1.3) \times 10^{16}\,$Hz). A decrease is then observed the following day (shown with a marker facecolor in white).\par

Owing to the dependence of the peak frequencies on used the fitting function, we also determine the synchrotron peak frequency following the phenomenological description of~\cite{blazar_seq}. We obtained values of $\nu_p$ higher by a factor of 2-3 compared to the log-parabola fit. Since the peak is not well covered for \textit{IXPE~1} and \textit{IXPE~2}, and rather flat for \textit{IXPE~3}, these model-dependent differences are expected. The clear trend of a synchrotron peak shifting towards higher values for \textit{IXPE~3} is still present.\par

The obtained spectral parameters in X-rays and VHE gamma rays are listed in Tab.~\ref{tab:IXPE1}. As for Fig.~\ref{fig:SED}, the MAGIC spectral fits are performed after averaging all nights within the \textit{IXPE} windows. For all \textit{IXPE} epochs, the MAGIC data show a preference for a log-parabola model (see Eq.\ref{eq:log_par}) over a simple power-law function. The preference is above $3\sigma$ for \textit{IXPE~1} and at the level of $2\sigma$ for \textit{IXPE~2}. We do not observe significant variability of the curvature parameter $\beta$, which stays consistent with $\beta=0.50$. Thus, throughout this work, the MAGIC spectra simultaneous to the \textit{IXPE} observations are fitted using a log-parabola model using a fixed curvature $\beta=0.50$. This choice removes any correlation between $\alpha$ and $\beta$ (see Eq.~\ref{eq:log_par}), providing a better assessment of the hardness evolution during the different epochs. The normalization energy is fixed to 300\,GeV. The resulting best fit spectral indices of MAGIC are shown in the first primary row of Tab.~\ref{tab:IXPE1}.\par

The \textit{Swift}-XRT spectra show a significant preference for a log-parabola model over a power-law. As in the MAGIC spectral study, the data are fitted using a log-parabola with fixed curvature in order to obtain a better characterization of the hardness evolution throughout the \textit{IXPE} epochs. We use here $\beta=0.29$, which is the average curvature over the campaign. The second primary row of Tab.~\ref{tab:IXPE1} presents the best fit parameters for each exposure simultaneous to \textit{IXPE} (the pivot energy of the log-parabola model is 1\,keV).\par 

Regarding \textit{XMM-Newton} and \textit{NuSTAR}, the spectral parameters are derived in the 0.6-10\,keV and 3-79\,keV bands, respectively. Similarly to the fits for MAGIC and \textit{Swift}-XRT, we fixed the curvature in the log-parabola model to $\beta=0.2$ for \textit{XMM-Newton} and to $\beta=0.45$ for \textit{NuSTAR}. For both instruments, the pivot energy is 1\,keV.\par

Overall, the spectral evolution is consistent with the typical harder-when-brighter trend found frequently in Mrk~421 \citep{2021MNRAS.504.1427A, 2021A&A...655A..89M}. At VHE, $\alpha$ during \textit{IXPE~2} is smaller compared to \textit{IXPE~1} ($\alpha=-2.30 \pm 0.08$ versus $\alpha=-2.64 \pm 0.06$ for \textit{IXPE 1}), while the emitted flux doubled. A similar behavior is found in X-rays with \textit{Swift}-XRT, \textit{XMM-Newton} and \textit{NuSTAR} data and confirmed by the visual trend in Fig.~\ref{fig:SED}. The spectral hardening is particularly evident between \textit{IXPE~2} and \textit{IXPE~3} when the X-ray polarization angle rotates. Both in \textit{Swift}-XRT and \textit{NuSTAR} the spectral parameter $\alpha$ hardens by $\approx0.3-0.4$ (see Tab.~\ref{tab:IXPE1}).\par

Most of the spectral variability in X-rays occurs on $\sim$daily timescale. The shorter timescales variability can be probed thanks to the multi-hour exposures from \textit{XMM-Newton} and \textit{NuSTAR}. Fig.~\ref{fig:xmm_lc_ixpe1} and Fig.~\ref{fig:xmm_lc_ixpe2} (Appendix~\ref{sec:appendix_xmm_light_curve}) show the 0.3-2\,keV the 2-10\,keV fluxes (binned in 500\,s) as well as the hardness ratio obtained during the observations of \textit{XMM-Newton}. The ratios do not reveal any prominent spectral evolution over $\sim$hour timescales for either days. The \textit{NuSTAR} analysis, however, reveals a moderate spectral change on $\sim$hour timescales, although spectral hysteresis behavior can be noticed. The more detailed analysis is presented in Sect.~\ref{sec:intranight_full_nustar}. \par

\subsection{Broadband evolution of the polarization degree between the IXPE epochs}
\label{sec:pol_deg_comp}

\begin{figure}[h!]
\centering
  \includegraphics[width=1\columnwidth]{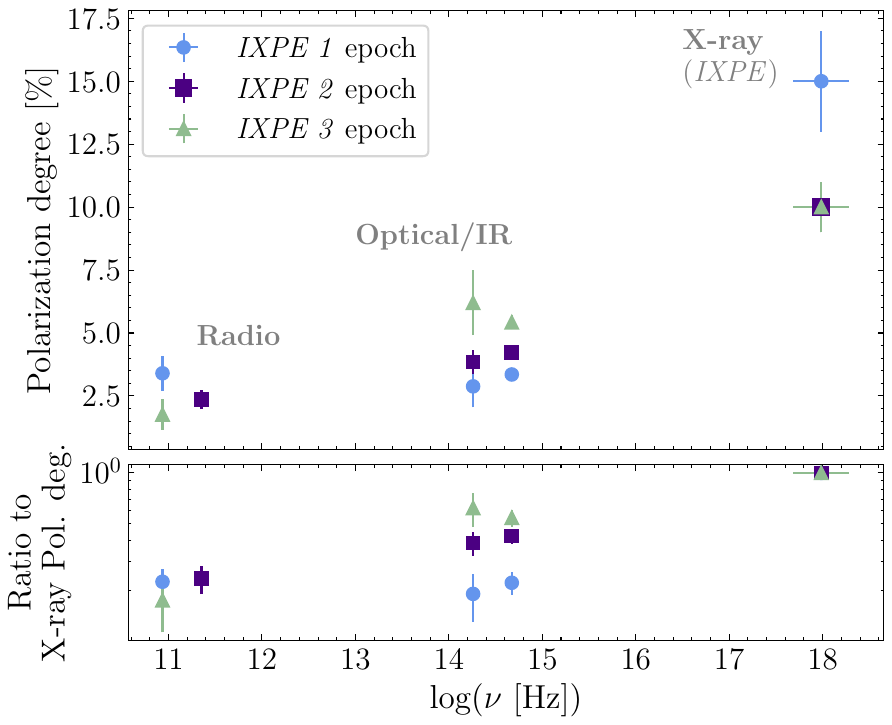}
  \caption{Top: Multiwavelength polarization degree as a function of frequency during all three \textit{IXPE} epochs. Bottom: Ratio of the frequency dependent polarization degree to the corresponding X-ray polarization degree.}
  \label{fig:pol_vs_freq}
\end{figure}

Fig.~\ref{fig:pol_vs_freq} summarizes the polarization degree as function of the frequency for all \textit{IXPE} observing epochs. The bottom panel shows the ratio to the X-ray polarization degree. For the optical/IR, we perform a weighted average of the measurements within the \textit{IXPE} observing windows. In the radio, we consider all measurements within the \textit{IXPE} windows as well as those that took place less than half a day before the start or after the end of the \textit{IXPE} observing times (i.e., all radio observations within MJD~59702.96 to MJD~59706.04, MJD~59733.99 to MJD~59736.94 and MJD~59736.90 to MJD~59739.88; May 3\textsuperscript{rd}-- 7\textsuperscript{th} 2022, June 3\textsuperscript{rd}-- 6\textsuperscript{th} 2022 and June 6\textsuperscript{th}-- 9\textsuperscript{th} 2022). This more relaxed simultaneity criteria allows one to include radio measurements for \textit{IXPE~2} and \textit{IXPE~3} epochs, which do not contain strictly simultaneous radio polarimetry coverage. We note that the variability of the radio polarization throughout this campaign is anyhow low and happens on timescales longer than 1\,day. Fig.~\ref{fig:pol_vs_freq} highlights the energy dependency of the polarization degree, with an evident increase in the X-ray band, as already reported by \citet{IXPE_Mrk421} and \citet{IXPE_Mrk501}, both in Mrk~421 and Mrk~501.\par

\begin{figure}[h!]
\centering
  \includegraphics[width=1.\columnwidth]{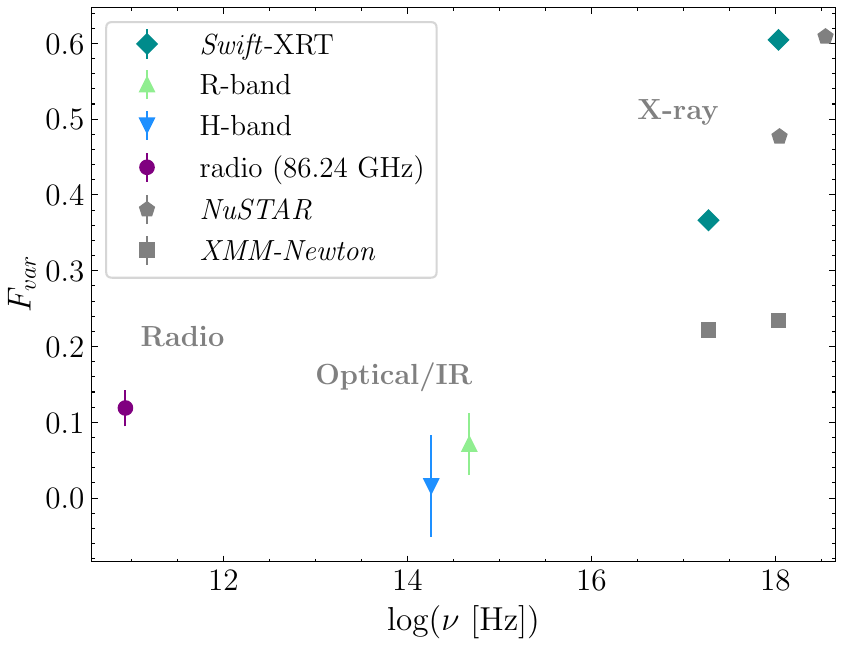}
  \caption{Fractional variability ($F_{var}$) as function of the frequency during the \textit{IXPE} epochs. $F_{var}$ is computed using all data from Fig.~\ref{fig:longterm_LC} that are within the \textit{IXPE} time windows. In the radio, we consider slightly relaxed simultaneity criteria and also include measurements that took place less than half a day before the start or after the end of the \textit{IXPE} observing times (see text for more details). Radio, optical/IR and \textit{Swift} data are daily binned. We include the $F_{var}$ from the \textit{NuSTAR} and \textit{XMM-Newton} multi-hour exposures using $\approx 30$\,min binning. The latter measurements are plotted in the grey since the two instruments did not gather data for all \textit{IXPE} epochs, which biases the comparison with other wavebands.}
  \label{fig:fvar_vs_freq}
\end{figure}

All epochs share the common characteristics of a significantly higher polarization in X-rays compared to lower frequencies. This highlights the value of combining X-ray and optical/radio polarization data. We do not find any significant correlation of the polarization degree with the flux or spectral hardness in the individual energy bands. On the other hand, the ratio between the optical/IR polarization degree and the one in the X-ray band is significantly lower during \textit{IXPE~1} than during \textit{IXPE~2} and \textit{IXPE~3} (bottom panel of Fig.~\ref{fig:pol_vs_freq}). \par

It is interesting to compare the broadband behavior of the polarization degree with the one of the fractional variability \citep[$F_{var}$;][]{2003MNRAS.345.1271V}. We compute $F_{var}$ using all observations inside the \textit{IXPE} windows, using the prescription of \citet{2008MNRAS.389.1427P} to estimate the corresponding uncertainties. The results are presented in Fig.~\ref{fig:fvar_vs_freq}. As for Fig.~\ref{fig:pol_vs_freq}, we consider in the radio a slightly relaxed simultaneity criteria and also include measurements that took place less than half a day before the start or after the end of the \textit{IXPE} observing times to compute $F_{var}$. The $F_{var}$ in the radio can only be computed with data from IRAM in the 86.24\,GHz band since it is the only one that has more than one measurement (that is the minimum requirement for a computation of $F_{var}$). In X-rays, we use \textit{Swift}-XRT fluxes binned observation-wise in the 0.3-2\,keV and 2-10\,keV ranges. We complement them with those from the \textit{XMM-Newton} (0.3-2\,keV and 2-10\,keV) and \textit{NuSTAR} (3-7\,keV and 7-30\,keV) long exposures. In the latter two cases, the results are plotted in grey markers to differentiate between them. Indeed, neither of the two instruments have simultaneous data for all \textit{IXPE} epochs (unlike \textit{Swift}-XRT), and given the generally stronger variability at those energies, this different temporal coverage biases the results and explains the discrepancy relative to the \textit{Swift}-XRT $F_{var}$.\par 

Similarly to the polarization degree, $F_{var}$ shows a significant increase in X-rays, while the optical and radio band are compatible within $1\sigma$. This trend, previously reported in Mrk~421 and other HSPs \citep{2015A&A...576A.126A, 2018A&A...611A..44P}, potentially suggests an underlying physical origin common to the one explaining the broadband behavior of the polarization degree. A discussion on this aspect is given in Sect.~\ref{sec:discussion}.

\begin{figure}[h!]
\centering
  \includegraphics[width=1.\columnwidth]{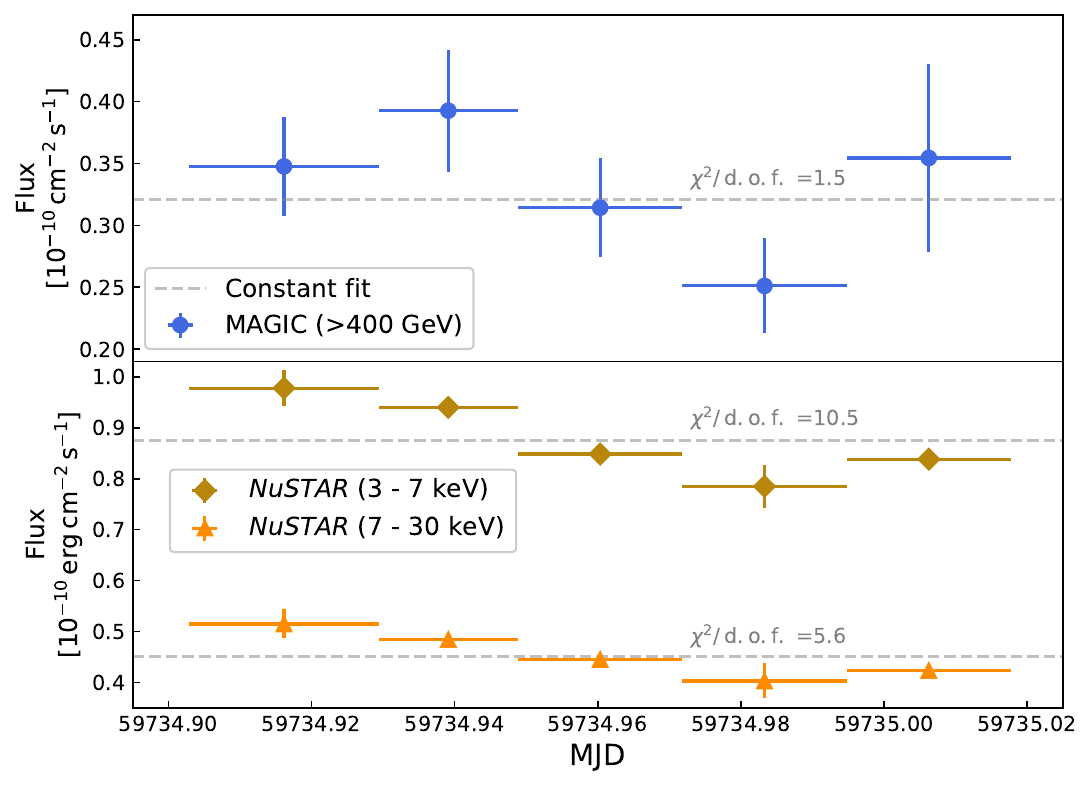}
  \caption{MAGIC and \textit{NuSTAR} intra-night light curves between 4\textsuperscript{th} June 2022 (MJD~59734) and 5\textsuperscript{th} June 2022 (MJD~59735), corresponding to the \textit{IXPE~2} epoch. Upper panel: Light curve above 400$\,$GeV obtained with MAGIC. A constant model fit is shown in dashed grey with the corresponding reduced $\chi^2$. Lower panel: Light curves for the 3-7$\,$keV and the 7-30$\,$keV bands taken by \textit{NuSTAR}, and constant fits for both. Fluxes from both instruments are computed in $\approx30$\,min time bins, except for the first bin that is $\approx40$\,min long due to a limited exposure of \textit{NuSTAR} around at the start of the MAGIC observation. }
  \label{fig:intranight}
\end{figure}
\begin{figure}[h!]
\centering
  \includegraphics[width=1.\columnwidth]{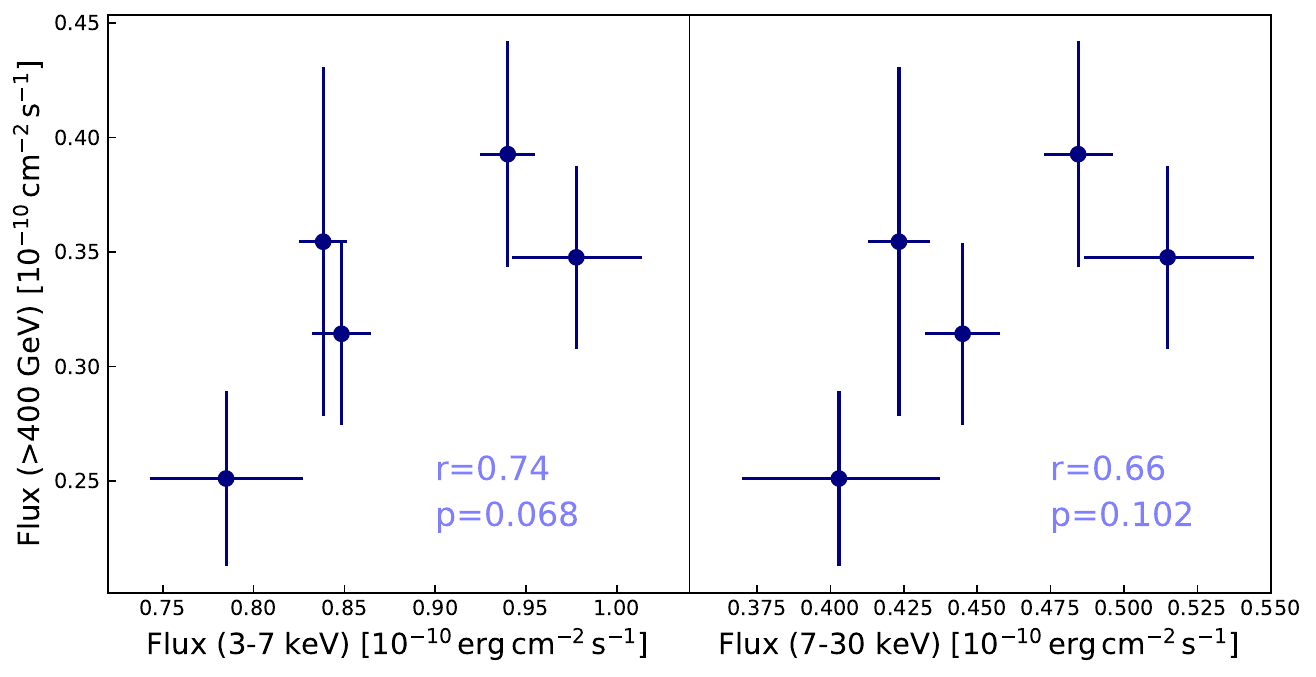}
  \caption{MAGIC flux versus \textit{NuSTAR} flux and quantification of the correlation during the \textit{IXPE~2} epoch. The MAGIC flux is computed above 400$\,$GeV while the X-ray flux is evaluated in two different energy bands: 3-7$\,$keV for the left panel and 7-30$\,$keV for the right panel. In each panel, the obtained Pearson's $r$ coefficient is indicated. We specify below the $p$-value describing the probability of obtaining the observed $r$ coefficient for two uncorrelated light curves. The latter $p$-value is estimated based on Monte Carlo toy simulations (see text for more details).}
  \label{fig:flux_flux}
\end{figure}

\begin{figure*}
\centering
  \includegraphics[width=2\columnwidth]{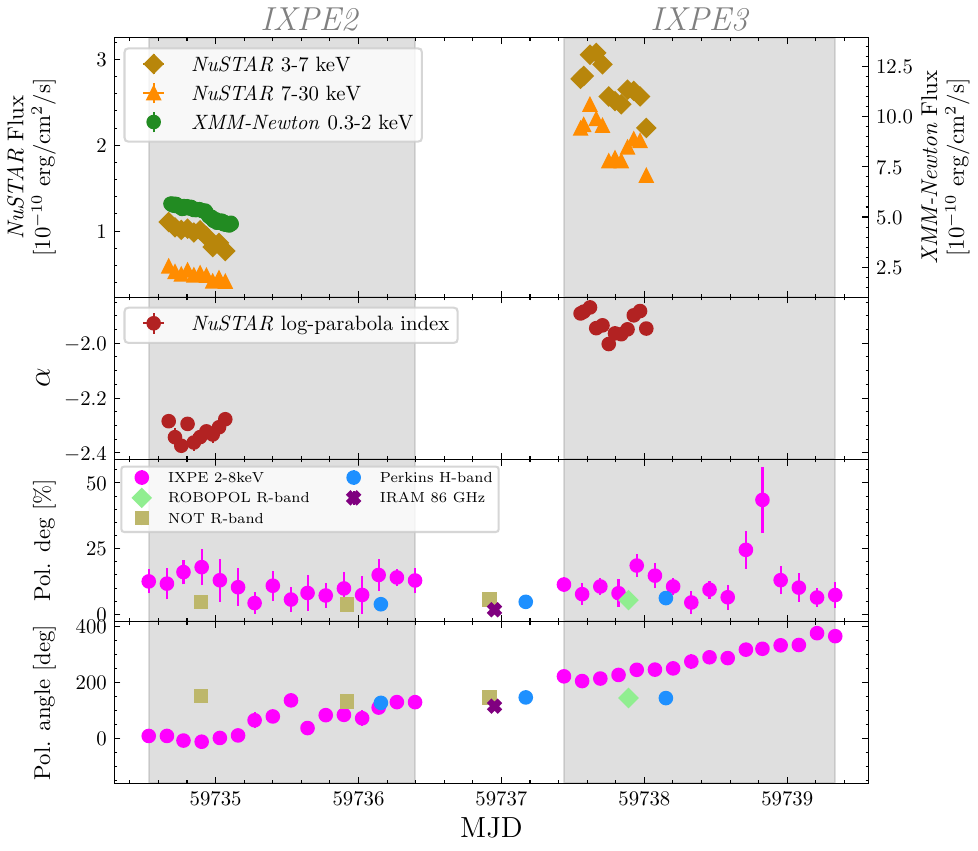}
  \caption{Zoom on the \textit{NuSTAR} light curves in the 3-7\,keV and 7-30\,keV bands during the \textit{IXPE~2} and \textit{IXPE~3} epochs. The top panel report the fluxes in 1\,hour bins. The second panel from the top is the $\alpha$ index evolution derived from fits of the \textit{NuSTAR} spectra. The last two panels show the simultaneous polarization degree and polarization angle in the X-ray band (\textit{IXPE}) and optical/radio bands.}
  \label{fig:intranight_nustar_ixpe}
\end{figure*}

\subsection{Intra-night MAGIC and \textit{NuSTAR} light curves during \textit{IXPE~2}}
\label{sec:intranight_ixpe2}

During the night of the 5\textsuperscript{th} to the 6\textsuperscript{th} of June 2022 (MJD~59734 to MJD~59735), MAGIC observations took place strictly simultaneously with \textit{NuSTAR}. The light curves obtained are shown in Fig.~\ref{fig:intranight}. The data are divided into bins of around 30$\,$min. Due to the otherwise limited exposure time by \textit{NuSTAR}, the first bin is extended to $\approx40$\,min. The upper panel shows the MAGIC fluxes, with an energy threshold of 400$\,$GeV. This minimum energy is slightly higher than in Fig.~\ref{fig:longterm_LC} since some of the time bins contain observations taken under a zenith distance of up to 60$^{\circ}$, which increases the energy threshold of the MAGIC stereo system. The \textit{NuSTAR} fluxes are extracted in the 3-7\,keV and 7-30\,keV bands.\par

No significant intra-night variability can be claimed for the MAGIC observations. On the other hand, \textit{NuSTAR} detects significant variability in both energy bands. By fitting the data with a constant model, the hypothesis of a non-variable emission is rejected at a significance above 5$\sigma$ for the 3-7$\,$keV band and above 3$\sigma$ for the 7-30$\,$keV band.\par

\begin{figure*}[h!]
\centering
  \includegraphics[width=2.\columnwidth]{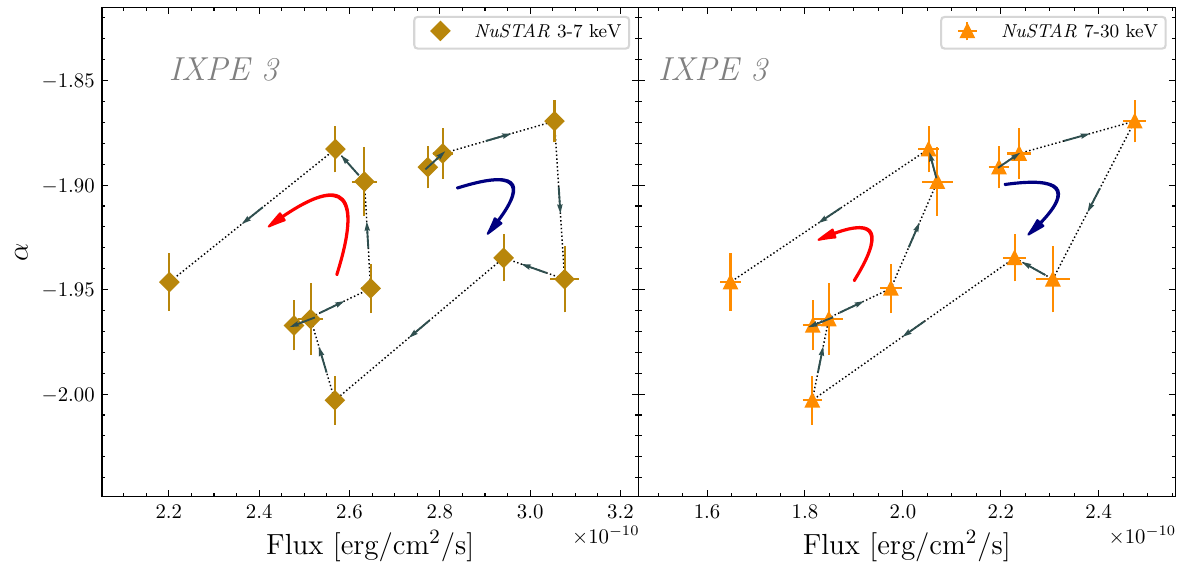}
  \caption{Log-parabola photon index $\alpha$ versus 3-7\,keV and 7-30\,keV flux as measured by \textit{NuSTAR} during the third \textit{IXPE} observation (\textit{IXPE~3} period). The data are 1\,hour binned and $\alpha$ is obtained by fitting a log-parabola function that has a fixed curvature parameter $\beta=0.45$. The grey arrows show the direction of time, and the blue and red arrows in the middle of the panels depict the clockwise and counter-clockwise directions observed in the data, respectively.}
  \label{fig:alpha_vs_flux_nustar}
\end{figure*}

The flux measured by MAGIC is plotted against the flux of both \textit{NuSTAR} energy bands in Fig.~\ref{fig:flux_flux}. The correlation coefficient between each pair of energy bands is given by the Pearson's $r$ coefficient. For the correlation between the 3-7$\,$keV and $>400\,$GeV flux a coefficient of $r=0.74$, and between 7-30$\,$keV and $>400\,$GeV of $r=0.66$ was found. Both cases suggest a light positive correlation. In order to evaluate the significance of the correlation, we use Monte Carlo simulated light curves. Each simulated flux point is produced by assuming a Gaussian distribution by taking the flux values of the actual data as a mean of the distribution and the uncertainty on the flux as the corresponding standard deviation. New light curves are then drawn and additionally the temporal information is shuffled in order to obtain pairs of \textit{realistic} uncorrelated light curves. We simulate $10^6$ pairs of light curves and derive $p$-values of the correlation coefficient $r$ of the data based on the distribution of the $r$ coefficients given by the simulations. We find a $p$-value of 0.068 (equivalent to $\approx 1.8\sigma$) between the 3-7$\,$keV and $>400\,$GeV bands and a $p$-value of 0.102 (equivalent to $\approx1.6\sigma$) for 7-30$\,$keV and $>400\,$GeV. Due to the relatively large statistical uncertainties in the VHE light curve, no significant correlation can be claimed and only an indication of correlation at best can be proposed.\par

In Sect.~\ref{sec:long_term_MWL}, we extend the search for correlation over longer timescale by including data from the entire MWL campaign between April 2022 to June 2022.

\subsection{Evidence of X-ray spectral hysteresis simultaneous to a polarization angle swing during \textit{IXPE~2} and IXPE~3}
\label{sec:intranight_full_nustar}

Using the multi-hour exposure of \textit{NuSTAR}, we investigate in detail the X-ray spectral evolution during the period where a polarization angle swing is detected by \textit{IXPE} in the X-rays (see Sect.~\ref{sec:IXPE_23_mwl} and Fig.~\ref{fig:longterm_LC}). Fig.~\ref{fig:intranight_nustar_ixpe} is a zoom around the polarization angle swing, showing the \textit{NuSTAR} measurements together with the polarization degree and angle in the radio/optical/IR. For \textit{IXPE}, the polarization degree and angle are binned in $\sim 3$\,hours.\par 

The top panel shows the \textit{NuSTAR} fluxes in the 3-7\,keV and 7-30\,keV bands, in 1\,hour time bins. Small variability is noted during the observation from the 5\textsuperscript{th} to the 6\textsuperscript{th} of June 2022 (MJD~59734 to MJD~59735, simultaneous to \textit{IXPE~2}), but more structured variability patterns can be seen during the observation simultaneous to \textit{IXPE~3}, between the 8\textsuperscript{th} and the 9\textsuperscript{th} of June 2022 (MJD~59737 to MJD~59738). In particular, the light curve displays two "humps" caused by two consecutive flux rise and decay phases, which thus reveal variability on $\sim$1\,hour timescale. \par 

The \textit{NuSTAR} spectra are fitted in the 3-30\,keV band adopting a log-parabola model (pivot energy fixed at 1\,keV). By fitting the spectra with a 1\,hour temporal binning, we find that the curvature parameter $\beta$ shows little variability throughout the observations. The derived $\beta$ values range from 0.27 to 0.57, but for each time bin they are within less than $\approx2\,\sigma$ from the weighted average over the two observations, which yields $\beta_{avg}=0.45$. Consequently, we perform a second series of fits with a 1\,hour binning after fixing $\beta=0.45$ to remove any correlation between $\alpha$ and $\beta$ in order to obtain a more straightforward assessment of the spectral hardness evolution. We stress that fixing $\beta=0.45$ does not significantly degrades the fit statistics (the beta-free spectral model is preferred at a significance below $2.5\,\sigma$ in each of the bins). The resulting index $\alpha$ as a function of time in 1\,hour bins is plotted in the second panel from the top in Fig.~\ref{fig:intranight_nustar_ixpe}.\par 

For the observation simultaneous to the \textit{IXPE~2} period (around MJD~59735 -- June 5\textsuperscript{th} 2022), we do not find strong spectral change. The index $\alpha$ varies by at most 5\% around a value of $\approx-2.35$, during a quasi-monotonic flux decay of $\approx 30\%$. We do not detect any significant correlation between $\alpha$ and flux, nor any spectral hysteresis pattern.\par 

Regarding the \textit{NuSTAR} observations simultaneous to the \textit{IXPE~3} period, a similar spectral variability amplitude is observed although hysteresis patterns can be seen when $\alpha$ is reported as function of the flux. Fig.~\ref{fig:alpha_vs_flux_nustar} shows the value of $\alpha$ versus the \mbox{3-7\,keV} and \mbox{7-30\,keV} fluxes in 1\,hour bins during \textit{IXPE~3}. The grey arrows indicate the direction of time. During the first part of the observation, the data points (both in the 3-7\,keV and 7-30\,keV bands) display a spectral hysteresis in a clockwise direction (i.e., decay phase has softer spectrum than in the rising phase). On the other hand, the second part of the observations exhibits a spectral hysteresis in counter clockwise direction (i.e., decay phase has a harder spectrum than in the rising phase). Spectral hysteresis, in both the clockwise and counter-clockwise direction, has been previously detected in Mrk~421 \citep{2003A&A...402..929B, 2004A&A...424..841R}. Nonetheless, it is the first time that two continuous clockwise and counter-clockwise rotations are detected over an hour timescale. A more detailed discussion of these results is given in Section~\ref{sec:discussion}.\par 

As unveiled by the bottom panels of Fig.~\ref{fig:intranight_nustar_ixpe}, no significant variability is observed in the polarization degree simultaneous to the \textit{NuSTAR} hysteresis patterns. Based on a constant fit, the data are consistent with a stable X-ray polarization hypothesis within $3\sigma$ (both for \textit{IXPE~2} and \textit{IXPE~3} periods). Regarding the X-ray polarization angle, the large angular swing mentioned before happens at a constant speed of $\sim80 ^\circ$/day despite the \textit{NuSTAR} flux and variability patterns discussed above.\\

\section{MWL evolution and correlation throughout the observing campaign}
\label{sec:long_term_MWL}
As commonly seen in HSPs such as Mrk~421, the flux (Fig.~\ref{fig:longterm_LC}) displays the strongest variability in the X-ray and VHE regimes. A noticeable feature in the MAGIC light curves is an enhanced VHE state period between MJD~59719 (20\textsuperscript{th} May 2022) and MJD~59723 (24\textsuperscript{th} May 2022). A peak flux of $\sim1.4$\,C.U. is measured on MJD~59722 (23\textsuperscript{rd} May 2022) in both the 0.2-1\,TeV and $>$1\,TeV bands (equivalent to $\approx3$ times the typical state). A simultaneous significant flux increase is noted in X-rays, as revealed by the \textit{Swift}-XRT light curves (third panel from the top). This high state also coincides with a hardening of both the VHE and X-ray spectrum, as illustrated by the hardness ratio plotted in the fourth panel from the top. This behavior, already seen within the \textit{IXPE} observing epochs in the earlier section, is consistent with the harder-when-brighter trend previously detected in Mrk\,421 \citep{2015A&A...576A.126A, 2021MNRAS.504.1427A, 2021A&A...655A..89M}. At lower energies, no simultaneous outburst is detected in the UV and optical (seventh and eighth panel from the top). On the other hand, it is interesting to remark that a RoboPol measurement (R-band) simultaneous to the peak activity at VHE on MJD~59722 (May 23\textsuperscript{th} 2022) shows a rotation of the polarization angle by about $60^\circ$ compared to an observation conducted a few days earlier ($\sim$MJD~59718 -- May 19\textsuperscript{th} 2022). Such a swing of the polarization angle of comparable amplitude and on similar timescales (i.e., $\sim$daily timescale) was reported by \citet{2021Galax...9...27M} in 2017, also for Mrk\,421. The sparse sampling of the RoboPol light curve prevents, however, a strong claim on the association of the optical polarization angle rotation with the VHE/X-ray flare. Besides the enhanced state around MJD~59722 (May 23\textsuperscript{th} 2022), the emission in the X-ray and VHE bands along the campaigns remains comparable to the quiescent activity. In fact, during previous outbursts, the VHE and X-ray fluxes were more than an order of magnitude higher than the average value from the campaign discussed in this work \citep{2020ApJS..248...29A, 2020ApJ...890...97A}. \par 

In the 0.3-300\,GeV band, the \textit{Fermi}-LAT light curve exhibits a flux variability by a factor $\sim3$ around an average state of $\sim 8\times 10^{-8}$\,cm$^{-2}$\,s$^{-1}$, which is close to the typical flux level for Mrk\,421 \citep{2015A&A...576A.126A}. As for the spectral evolution, no significant variability of the \textit{Fermi}-LAT power-law index is detected.\par 

In the UV band, despite moderate variability, the \textit{Swift}-UVOT fluxes display an interesting quasi monotonic increase starting from $\sim$MJD~59710 (11\textsuperscript{th} May 2022) to $\sim$MJD~59760 (30\textsuperscript{th} June 2022). The highest UV state is registered on MJD~59753 (23\textsuperscript{rd} June 2022), and slightly more than twice the minimum state is measured on MJD~59717 (18\textsuperscript{th} May 2022). The R-band measurements show a similar increasing trend over this period. Over the same period, the X-ray band shows an opposite evolution with an overall decay of the 0.3-2\,keV and 2-10\,keV fluxes. The latter behavior is accompanied by a simultaneous drop of the X-ray hardness ratio (see the fourth panel from the top in Fig.~\ref{fig:longterm_LC}). Such a behavior points towards an anti-correlation between the X-ray and UV bands, possibly caused by a shift of the entire synchrotron component to lower frequencies. The quantification of the anti-correlation significance is performed in Sect.~\ref{sec:anti_corr_uv_x_ray_analysis}.\par 

\subsection{VHE/X-ray correlation over the entire campaign} 

In Section~\ref{sec:intranight_ixpe2}, we reported an indication of positive correlation between the MAGIC and \textit{NuSTAR} fluxes during the \textit{IXPE~2} observations. The low significance (estimated around $2\sigma$ between the 3-7\,keV and $>400$\,GeV bands) is partly due to the relatively large uncertainties on the VHE gamma-ray fluxes measured in these short timescales. In this section, we extend the VHE vs X-ray correlation study over the entire campaign by making use of the MAGIC and \textit{Swift}-XRT measurements. We correlate the daily binned MAGIC fluxes (in the 0.2-1\,TeV and $>1$\,TeV bands) with the \textit{Swift}-XRT fluxes  binned observation-wise (0.3-2\,keV and 2-10\,keV bands), and compute the discrete correlation coefficient \citep[DCF;][]{1988ApJ...333..646E} in a series of 2-day binned time lags. The significance of the DCF is estimated based on Monte Carlo simulations. The simulations were performed in a similar fashioned to what is described in \citet{2021A&A...655A..89M}. We summarize below the procedure.\par

The significance bands are obtained by first simulating a large number ($10^4$) of \textit{uncorrelated} light curves for each of the energy bands considered. The light curves are simulated using the prescription from \citet{2013MNRAS.433..907E} in order to preserve the probability distribution function of the observed fluxes. Furthermore, the simulated light curves are produced by assuming a power spectral density (PSD) function that follows a power-law model. The slopes of the PSD models in X-rays are directly taken from \citet{2021A&A...655A..89M}, being $-1.45$ for the 0.3-2\,keV band and $-1.3$ for the 2-10\,keV band. These slopes (derived with \textit{Swift}-XRT data in 2016-2017 that cover a longer time span than the one considered in this work) are found to be in agreement with the 2022 observations and thus represent a good proxy to estimate the significance. Regarding the simulations of VHE light curves, it is not possible to directly extract the PSD slope in a reliable manner using the MAGIC data of this work due to the relatively sparse sampling. We therefore adopt the PSD slope of $-1.3$ that was reported by \citep{2015A&A...576A.126A} using Whipple observations during a campaign organized in 2009. The fake light curves are generated with a temporal resolution matching the typical exposure time of the observations, and the same temporal sampling as the data is then applied to the simulations. Finally, we compute the DCF as a function of time lag for all pairs of simulated light curves. The $2\sigma$, $3\sigma$ and $4\sigma$ confidence bands are derived from the distribution of the simulated DCF values in each time lag bin.\par

\begin{figure}
\centering
  \includegraphics[width=1.\columnwidth]{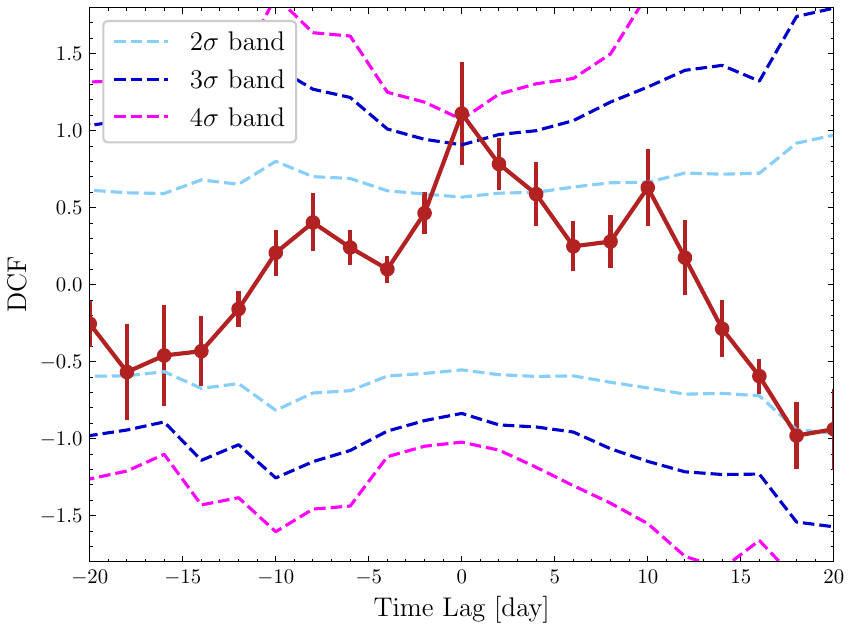}
  \caption{Discrete correlation function DCF computed for the MAGIC $0.2-1$\,TeV and \textit{Swift}-XRT 2-10$\,$keV light curves between MJD~59700 (May 1\textsuperscript{st} 2022) and MJD~59740 (June 10\textsuperscript{th} 2022) with a time-lag binning of 2\,days. The red points are the obtained DCF values and their uncertainties. The light blue, dark blue and pink dashed lines show the $2\sigma$, $3\sigma$ and $4\sigma$ significance bands, respectively (see text for more details).}
  \label{fig:DCF_magic_LE_Xray_he}
\end{figure}
\begin{figure}
\centering
  \includegraphics[width=1.\columnwidth]{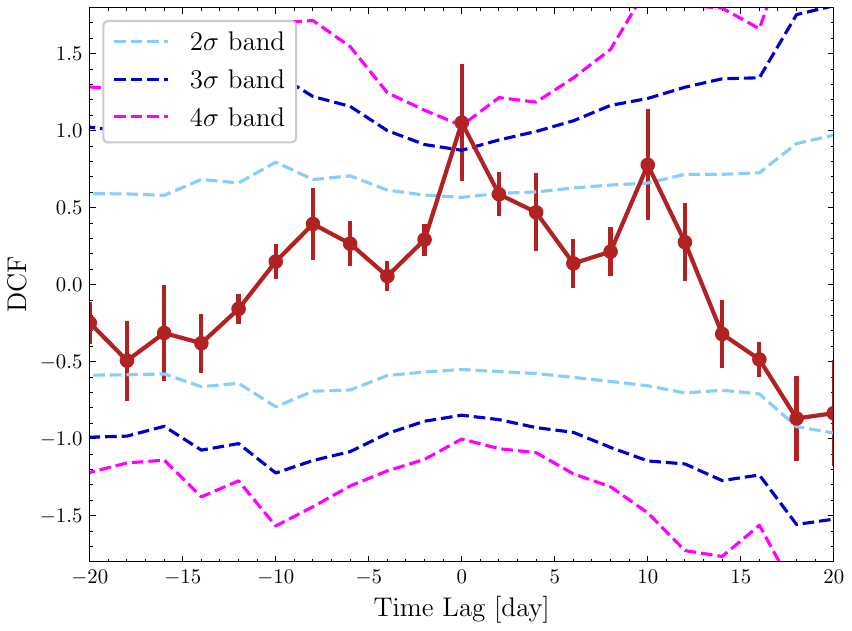}
  \caption{Discrete correlation function DCF computed for the MAGIC $>1$\,TeV and \textit{Swift}-XRT 2-10$\,$keV light curves between MJD~59700 (May 1\textsuperscript{st} 2022) and MJD~59740 (June 10\textsuperscript{th} 2022) with a time-lag binning of 2\,days. The red points are the obtained DCF values and their uncertainties. The light blue, dark blue and pink dashed lines show the $2\sigma$, $3\sigma$ and $4\sigma$ significance bands, respectively (see text for more details).}
  \label{fig:DCF_magic_HE_Xray_he}
\end{figure}

Fig.~\ref{fig:DCF_magic_LE_Xray_he} and Fig.~\ref{fig:DCF_magic_HE_Xray_he} show the DCF obtained from MAGIC 0.2-1\,TeV versus \textit{Swift}-XRT 2-10\,keV and MAGIC $>1$\,TeV versus \textit{Swift}-XRT 2-10\,keV, respectively. The dashed lines depict the $2\sigma$ (light blue), $3\sigma$ (dark blue), $4\sigma$ (magenta) confidence bands. A positive correlation can be seen at zero time lag with a significance of $4\sigma$, further strengthening the  reported in Section~\ref{sec:intranight_ixpe2}. As for the correlation of the MAGIC fluxes with the \mbox{0.3-2\,keV} band, the significance is somewhat lower, around $3\sigma$. The results are shown in Fig.~\ref{fig:DCF_magic_LE_Xray_le} and Fig.~\ref{fig:DCF_magic_HE_Xray_le} in Appendix~\ref{sec:appendix_vhe_vs_xray}. This suggests that the \mbox{2-10\,keV} flux is more closely related to the VHE flux compared to the \mbox{0.3-2\,keV} band during this period of time.

\subsection{Investigation of the UV/optical versus X-ray anti-correlation}
 \label{sec:anti_corr_uv_x_ray_analysis}

Fig.~\ref{fig:longterm_LC} suggests an anti-correlation between the UV and X-ray fluxes between MJD~59710 (May 11\textsuperscript{th} 2022) and MJD~59760 (June 30\textsuperscript{th} 2022). We quantify this trend by computing the DCF between the \textit{Swift}-XRT data (using both the 0.3-2\,keV and 2-10\,keV fluxes) and the \textit{Swift}-UVOT W1 measurements. For simplicity, only the data in the \textit{Swift}-UVOT W1 band are considered for this correlation study. In fact, the fluxes in the M2 and W2 \textit{Swift}-UVOT filters give very similar results, which is expected given their proximity in frequency with W1. The resulting plots are shown in Appendix~\ref{sec:appendix_uv_vs_xray} in Fig.~\ref{fig:DCF_UV_Xray_le} and Fig.~\ref{fig:DCF_UV_Xray_he} for the 0.3-2\,keV and 2-10\,keV bands, respectively. The significance bands are obtained with the exact same method described in the previous section. The PSD slopes are taken from \citet{2021A&A...655A..89M}, i.e., $-1.45$ for \textit{Swift}-UVOT W1 and \textit{Swift}-XRT 0.3-2\,keV and $-1.3$ for \textit{Swift}-XRT 2-10\,keV. We find that the significance of the anti-correlation observed in the data is at the level of $2-3\sigma$, and can only be considered as a marginal evidence. The significance is marginally higher in the \textit{Swift}-UVOT W1 versus \textit{Swift}-XRT 2-10\,keV case than in the \textit{Swift}-UVOT W1 versus \textit{Swift}-XRT 0.3-2\,keV case. The peak at a positive time lag of $\sim$16 days in both figures, can be considered an artifact resulting from the sampling and short overall time period.\par 

We repeated the above exercise after including \textit{Swift} data from the entire MWL campaign (i.e. from MJD~59695 to MJD~59760; April 26\textsuperscript{th} 2022 to June 30\textsuperscript{th} 2022). The results - shown in Fig.~\ref{fig:DCF_UV_Xray_le_full_camp} and Fig.~\ref{fig:DCF_UV_Xray_he_full_camp} from Appendix~\ref{sec:appendix_uv_vs_xray} - reveal a decrease in the significance below $2\sigma$. The marginal evidence of anti-correlation is thus only observed over a 1.5\,months period between $\sim$MJD~59710 (May 11\textsuperscript{th} 2022) and $\sim$MJD~59760 (June 30\textsuperscript{th} 2022). \par 

This is the third time that an indication of anti-correlation between UV and X-ray fluxes is reported in Mrk~421. The first two hints were observed during MWL campaigns organized during 2009 \citep{2015A&A...576A.126A} and 2017 \citep{2021A&A...655A..89M}, and were also happening over $\sim$monthly timescale. These repeating trends point towards some physical connection between the UV and X-ray emitting regions, which is particularly relevant in the context of the recent \textit{IXPE} results that suggest energy stratified emitting regions.\par 

The anti-correlation is not significantly detected during the first part of the 2022 campaign, which might be explained by a low variability. Alternatively, the physical mechanism responsible for the anti-correlation may only take place temporarily. \citet{2021A&A...655A..89M} investigated the anti-correlation between X-ray and UV as well as X-ray and optical over several months. They also found that such a trend became significant on $\sim$monthly timescales, possibly indicating that it is not a permanent feature of Mrk~421.\par

\subsection{Optical polarization evolution throughout the entire campaign}
 \label{sec:optical_pol_evol}

\begin{figure}
\centering
  \includegraphics[width=1.\columnwidth]{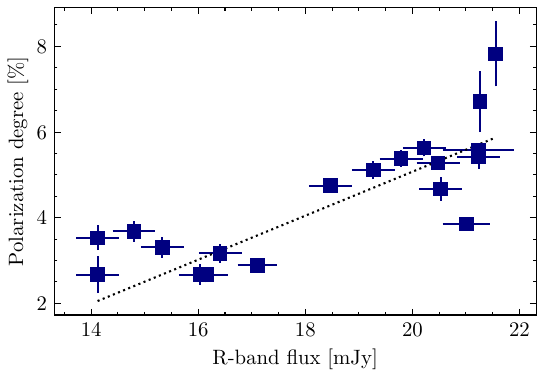}
  \caption{Correlation between the polarization degree and flux in the R-band over the entire campaign. The black dotted line is a linear fit, yielding a best-fit slope of $a=0.52\pm0.09$. The Pearson's $r$ of the correlation is $r=0.8\pm0.1$. The associated p-value is $p_{value}=1\times 10^{-5}$.}
  \label{fig:rband_vs_poldeg}
\end{figure}

The R-band flux, which is close to the UV in frequency, also displays an increase throughout the campaign, in particular during the second part (between MJD~59710 and MJD~59760; May 11\textsuperscript{th} 2022 to June 30\textsuperscript{th} 2022), corroborating the anti-correlation hint derived with the \textit{Swift}-UVW1 measurements. The R-band data are unfortunately too sparse to properly quantify the trend in the latter waveband. The rise of the optical flux seems to be accompanied by an increase of the polarization degree. In Fig.~\ref{fig:rband_vs_poldeg}, we present the correlation between the polarization degree and flux using strictly simultaneous R-band measurements. We consider all data from RoboPol, NOT and T90 along the campaign. The KANATA measurements are discarded because of their very large flux uncertainties, in comparison to the measurements from the other instruments. We stress that the data mostly cover the MJD~59710 to MJD~59760 period (i.e., during the UV/X-ray anti-correlation hint period; May 11\textsuperscript{th} 2022 to June 30\textsuperscript{th} 2022), except for a single NOT measurement that took place before (on MJD~59703 -- May 4\textsuperscript{th} 2022). We find a positive correlation with a Pearson's $r$ of $r=0.8\pm0.1$. Using the same method as in Sect.~\ref{sec:intranight_ixpe2}, we estimated an associated \mbox{p-value} of $p_{value}=1\times 10^{-5}$, corresponding to a correlation significance of $\approx4\sigma$. By fitting a linear function (see black dotted line), the slope of the correlation is $a=0.51\pm0.09$. The same results are derived if one considers data between MJD~59710 to MJD~59760 (i.e., after removing the NOT measurement on MJD~59703).\par 

Overall, the combination of the $\sim$monthly timescales UV/X-ray anti-correlation and the rise of the R-band polarization degree observed over the similar timescales potentially implies a general change in the physical properties of the source. The interpretation of this observation is given in Sect.~\ref{sec:discussion}.

\section{Discussion and Summary} \label{sec:discussion}

This work reports on an extensive MWL campaign on Mrk~421 organized in 2022 from radio to VHE gamma rays, including, for the first time, a simultaneous characterization of the X-ray polarization behavior. The VHE observations were carried out by the MAGIC telescopes, and are accompanied by observations from \textit{Fermi}-LAT, \textit{NuSTAR}, \textit{XMM-Newton}, \textit{Swift} as well as multiple instruments covering the optical to radio bands.\par 

During the first \textit{IXPE} observation in May 2022 (\textit{IXPE~1}), the daily coverage from MAGIC reveals a low emission state at VHE ($\approx25\%$ of the Crab Nebula in the 0.2-1\,TeV band) without any significant variability on either daily and hour timescales. Moderate daily variability is noted in the X-ray band, which reveals an emission state lower than the average activity of Mrk~421 \citep{abdo:2011}. The optical/UV and MeV-GeV fluxes remain close to the typical activity. As for the broadband polarization characteristics, the polarization degree is significantly stronger in X-rays than at lower frequencies. It illustrates the importance of combining X-ray and optical/radio polarization data. As discussed in \citet{IXPE_Mrk421} and \citet{IXPE_Mrk501}, those results are in line with an energy stratified jet, where the most energetic particles (emitting X-ray photons) are located in smaller regions that possess a more ordered magnetic field, close to the acceleration site. The energy dependency and the slow variability of the polarization degree strongly points towards a shock acceleration scenario. Electrons subsequently cool, and diffuse in larger regions where the field is more turbulent to further emit from optical to radio frequencies. During \textit{IXPE~1}, there is no significant variation in the polarization angle~\citep{IXPE_Mrk421} at any energies. In particular, the X-ray polarization angle is compatible with the one measured in the optical and radio. \par 

In June 2022, the \textit{IXPE~2} and \textit{3} epochs are also characterized by a constant X-ray polarization degree that is significantly higher compared to lower frequencies. Such a general broadband feature of the polarization degree shares some similarities with the variability strength (quantified with the fractional variability $F_{var}$), which also shows an increase with energy. The $F_{var}$ during the \textit{IXPE} exposures is indeed significantly higher in the X-ray band compared to the optical/radio data. The latter behavior may partially be caused by an X-ray emission dominated by (a single or a few) compact regions whose temporary appearance within the jet drives the observed variability, while emission at lower frequencies receives simultaneous contributions from several broader regions that decreases the overall variability. Such a scenario corroborates the energy stratification of the jet implied by the energy dependency of the polarization strength.\par

While the \textit{IXPE~2} and \textit{3} epochs are consistent with a constant polarization degree, the polarization angle exhibits an evident rotation in X-rays during the latter two \textit{IXPE} exposures. The rotation proceeds at constant angular velocity \citep[see also][]{mrk421_ixpe_rotation} between the two epochs, hence highly suggesting a single rotation event observed during the two consecutive \textit{IXPE~2} and \textit{3} exposures. The optical and radio observations do not reveal a simultaneous angle rotation.\par 

We manage to characterize the VHE state only at the very beginning of the polarization angle rotation. During that time period, we find a VHE emission state higher (and the spectrum is harder) than during \textit{IXPE~1}, although comparable to the average one for the source \citep[$\approx50\%$ of the Crab Nebula in the 0.2-1\,TeV band][]{abdo:2011}. Starting from the second half of the \textit{IXPE~2} epoch and during \textit{IXPE~3}, the activity in X-rays increases and hardens significantly simultaneously with the angle rotation. The emission reaches a maximum well above the Mrk~421 quiescent state. The VHE gamma rays usually show a strong correlation with X-rays, especially during X-ray flaring activities, as observed during \textit{IXPE~3} \citep[see e.g.,][]{2020ApJS..248...29A}, but the lack of simultaneous observations with MAGIC does not allow us to evaluate this characteristic during this specific flaring activity in June 2022.\par 

Previous campaigns on LSP and ISP objects have shown that rotations of the polarization angle in the optical can be associated with flares~\citep{2017A&A...603A..29A,2018A&A...619A..45M,2010Natur.463..919A, 2019AJ....157...95G, 2015ApJ...809..130C, 2008Natur.452..966M}. In LSPs and ISPs, the synchrotron peak is located around the optical band, while in HSPs (as Mrk~421) it is located in the X-ray regime. One would thus naively expect that X-ray flares in HSPs can similarly be associated with X-ray polarization angle swings. Even if the enhanced X-ray state during \textit{IXPE~2} and \textit{3} remains below previous notable outbursts of Mrk~421 \citep[see for instance the March 2010 flare reported in ][]{2015A&A...578A..22A}, the evident X-ray flux rise and hardening, temporally coincident with a swing of the polarization angle swing, may share a common physical origin as angular swings observed in lower synchroton peaked blazars. \par 

The absence of a simultaneous polarization angle swing in the optical/IR and radio may be explained by the following scenario: the smaller region radiating the X-ray photons (where the $B$ field is more ordered) is streaming down the jet following helical field lines, leading to an apparent rotation of the polarization angle \citep{mrk421_ixpe_rotation}, while at lower frequencies, the emitting regions are larger and does not closely follow a helical path as for the X-ray region. \par

\begin{figure}
\centering
  \includegraphics[width=1.\columnwidth]{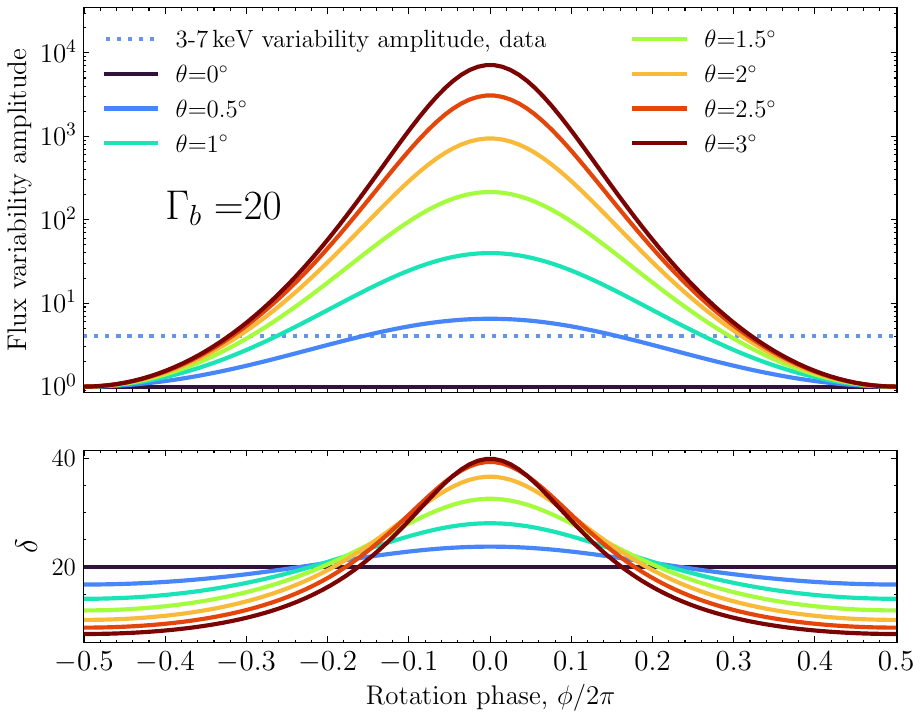}
  \caption{Top panel: Flux variability amplitude caused by the evolution of the Doppler factor $\delta$ when an emitting zone is travelling downstream along a helical path. The variability amplitude is plotted as function of the phase of the spiral rotation. The horizontal dotted blue line gives the observed variability amplitude in the 3-7\,keV band. Bottom panel: The corresponding Doppler factor $\delta$ as a function of the spiral rotation phase. The curves are produced for different jet axis angles ($\theta$) relative to the line of sight. The zone is assumed to move with a Lorentz factor $\Gamma_b=20$ in a helical field with a pitch angle of $2.9^\circ$ (see text for more details).}
  \label{fig:rotation_var_ampl}
\end{figure}

The movement of a compact region through a helical path inside the jet induces changes in the Doppler factor, which then lead to significant observed flux variability. We thus investigate if such a scenario, proposed to explain the angle rotation, is (roughly) consistent with the observed variability amplitude in the X-rays (that is the energy range with the best temporal coverage during the rotation). The viewing angle $\psi$ of a region streaming down an helical path is given by \citep[see e.g.][]{2013ApJ...768...40L}:
\begin{equation}
\psi = \arccos{ \left[ \cos{\theta} \cos{\zeta} +\sin{\theta} \sin{\zeta} \cos{\phi} \right] }
\end{equation} 

where $\theta$ is the jet axis angle to the line of sight, $\zeta$ the pitch angle of the helical field and $\phi$ the phase of the spiral rotation. If the region moves at a Lorentz factor of $\Gamma_b$, the associated Doppler factor is $\delta = \left[ \Gamma_b (1-\beta \cos{\psi} \right]^{-1}$. In the observer's frame, the intrinsic flux $F_{intr}$ transforms as \citep{1979rpa..book.....R}:

\begin{equation}
F_{obs} = \delta^{3+p} F_{intr}
\end{equation} 

where $p$ is the photon index, which we found to be around $-2.4$ in the 3-7\,keV band of \textit{NuSTAR} during the rotation. According to \citet{mrk421_ixpe_rotation}, the rotation rate can be reproduced if the emitting feature travels with a velocity component parallel to the jet axis of
0.9975$c$ and a transverse component of 0.05$c$. Based on this estimation, the corresponding pitch angle of the helical field is $\approx 2.9^\circ$. Assuming a typical Lorentz factor of $\Gamma_b=20$, the expected flux variability amplitude solely introduced by an evolution of the Doppler factor due to the movement on a helical path is plotted in the top panel Fig.~\ref{fig:rotation_var_ampl}. The variability is plotted as function of the phase of the rotation. The curves are plotted for a set of $\theta$ ranging from $0^\circ$ to $3^\circ$, which is typical for blazars. The variability amplitude is strongly dependent on the jet viewing axis, being a few orders of magnitude high for $\theta=3^\circ$. The horizontal blue dotted line displays the observed flux amplitude in the 3-7\,keV band, which can be explained by the change of Doppler factor if $\theta \approx 0.5^\circ$. The low apparent speed of radio knots in Mrk~421 suggest that very long baseline radio observations mostly probe the sheath of the jet instead of the central part \citep[i.e. the spine, see e.g.][]{2005A&A...432..401G,2022ApJS..260...12W}. We note however that \citet{2022ApJS..260...12W} estimated $\theta\sim1^\circ$, being rather consistent with Fig.~\ref{fig:rotation_var_ampl}. One concludes that, assuming relatively standard parameters, the observed flux changes are not in contradiction with the variability caused by the evolution of the Doppler factor (as the zone travels on helical field lines). It is important to note that the flux variability is also likely affected by acceleration and cooling processes, as suggested by the spectral changes observed on $\sim$\,day-hour timescale in the \textit{NuSTAR} data during \textit{IXPE~2~\rm{and}~3}. And hence the MWL data tells us that the changes in the $\delta$ cannot be the only reason for the observed flux variability.\par

The observations by \textit{NuSTAR} simultaneous to the polarization angle swing during \textit{IXPE~3} unveil two contiguous spectral hysteresis loops in opposite directions over a single exposure (see Fig.~\ref{fig:alpha_vs_flux_nustar}). The first loop, in a clockwise direction, is likely the signature of synchrotron cooling causing a delay of the low-energy X-ray photons with respect to the high-energy ones (soft lag). The subsequent counter-clockwise loop indicates a delay of the high-energy X-ray photons compared to the low-energy ones (hard lag), suggesting a system observed at energies for which acceleration timescale is comparable to the cooling timescale, $t'_{acc} \approx t'_{cool, synch}$ \citep{1998A&A...333..452K}.\par

Within a framework of shock acceleration, as suggested by the multi-band polarization properties, the acceleration timescale in the co-moving frame (in what follows, primed quantities are in the co-moving frame and unprimed quantities refer to the observer's frame) of an electron with Lorentz factor $\gamma'$ can be approximated as follows \citep{1983RPPh...46..973D, 1987PhR...154....1B, 2000ApJ...536..299K}:

\begin{equation}
    t'_{acc} = \frac{20 \lambda(\gamma') c }{3 u^2_s} 
\end{equation}
where $\lambda(\gamma') = \frac{\xi \gamma' m_e c^2}{eB'}$ is the mean free path of electrons, parameterized as a fraction $\xi$ of the Larmor radius. The parameter $\xi$, sometimes dubbed as the \textit{gyro-factor}, is a parameter related to the efficiency in the  acceleration of the high-energy particle population, and is always $\geq1$. In the so-called \textit{Bohm} limit, the acceleration is the most efficient because it occurs over a mean free path similar to the Larmor radius, and $\xi=1$. Within this framework, the acceleration efficiency is proportional to $\xi^{-1}$, and $\xi>1$ indicates an acceleration efficiency lower than that in the \textit{Bohm} limit. $B'$ is the magnetic field inside the emitting region and $u_s$ the speed of the shock, which we assume to be relativistic, $u_s \sim c$. The synchrotron cooling time is given by:

\begin{equation}
    \label{eq:synch_cooling_formula}
    t'_{cool, synch} = \frac{3 m_e c}{4 \sigma_T U'_{B} \gamma'} = \frac{6 \pi m_e c}{\sigma_T B'^2 \gamma'} 
\end{equation}
where $\sigma_T$ is the Thomson cross section, and $U'_{B}=B'^2/8\pi$ the magnetic field energy density. By expressing the latter timescales in terms of the observed photon energy, and considering that electrons emit most of their synchrotron photons at an observed frequency of $\nu \approx 3.7 \times 10^6 \frac{\gamma'^2 B' \delta}{1+z} $, where $\delta$ is the Doppler factor of the emitting region, one finds that the ratio $t_{acc}/t_{cool, synch}$ is in fact independent of $B'$ \citep{2002ApJ...572..762Z}:
\begin{equation}
\label{eq:ratio_timescale}
    \frac{t_{acc}}{t_{cool, synch}}(E) = 3.17 \times 10^{-5} (1+z) \xi \delta^{-1} E \quad {\rm s}
\end{equation}
where $E$ is the photon energy in keV units. The counter-clockwise loop observed by \textit{NuSTAR} implies $t_{acc}/t_{cool, synch} \approx 1$ at $E\approx10$\,keV, which is the characteristic energy probed by \textit{NuSTAR}. Assuming a typical $\delta=30$ for Mrk~421, one thus derives $\xi \approx 8 \times 10^4$ for the second part of the \textit{NuSTAR} observation during the \textit{IXPE~3} epoch.\par 

On the other hand, the first part of the \textit{NuSTAR} observation in the \textit{IXPE~3} epoch, where a clockwise loop is observed, suggests a regime in which \mbox{$t_{acc}/t_{cool, synch}<<1$} since synchrotron cooling is likely the driver of soft lags. The acceleration must take place in a significantly more effective manner. During this part of the observation, $\xi$ must therefore be at least an order of magnitude smaller, $\xi \lesssim 8 \times 10^3$. While the range of values we derive for $\xi$ stay within the estimates of \citet{2017MNRAS.464.4875B}, where it is discussed in a broader theoretical context, the consecutive clockwise and counter-clockwise loops during \textit{IXPE~3} imply an increase of the \textit{gyro-factor} $\xi$ of at least one order of magnitude over $\sim$hour timescales.\par

The above calculations and estimations of $\xi$ do not consider IC cooling. We verified that such simplification is not significantly affecting our results. Using a SSC model \citep{1992ApJ...397L...5M, 1999ApJ...521..145M} that we constrain using the X-ray \& VHE spectra during the \textit{IXPE} epochs, we estimate that the IC cooling timescale is longer than the synchrotron cooling timescale, as one would anyhow expect from the lower luminosity of the IC bump, in comparison to that of the synchrotron bump. Our model in fact shows a synchrotron cooling timescale that is about twice shorter than the IC cooling. Hence, the synchrotron cooling is sufficient to estimate the dynamics of the electrons and Eq.~\ref{eq:ratio_timescale} remains a valid approximation to estimate $\xi$. A detailed description of the model and the computation is given in Appendix~\ref{sec:modelling_xray_vhe}.\par

The modelling performed in Appendix~\ref{sec:modelling_xray_vhe} constrains the magnetic field to be $B' \sim 0.04$\,G in the X-ray/VHE emitting region with a blob radius of $R'\approx2\times10^{16}$\,cm. Those values imply a synchrotron cooling time (Eq.~\ref{eq:synch_cooling_formula}) longer than the light-crossing time ($t'_{cr}=R'/c$) for electrons emitting up to $\approx10$\,keV, which is well within the \textit{NuSTAR} bandwidth. The modelling parameters are thus clearly in agreement with a \textit{NuSTAR} variability regulated by cooling (and/or acceleration) mechanisms, instead of light-crossing time effects, as suggested by the observed hysteresis loops. If the light travel time would be significantly longer than the cooling/acceleration timescale, the variability will be dominated via the former.\par

As a final consideration, we combine Eq.~\ref{eq:synch_cooling_formula} with the characteristic synchrotron frequency ($\nu \approx 3.7 \times 10^6 \frac{\gamma'^2 B' \delta}{1+z} $) to derive the expected cooling time scale in the observer's frame \citep{2002ApJ...572..762Z}:
\begin{equation}
t_{cool, synch}(E) = 3.04 \times 10^3 B'^{-3/2} (1+z)^{1/2} \delta^{-1/2} E^{-1/2} \quad {\rm s}
\end{equation}
The parameters from the modelling in Appendix~\ref{sec:modelling_xray_vhe} gives $t_{cool, synch}(E=\rm{3\,keV}) \approx 11$\,hr, and $t_{cool, synch}(E=\rm{10\,keV}) \approx 6$\,hr, which is again well consistent with the flux doubling/halving timescale derived by the \textit{NuSTAR} data.

Within the \textit{IXPE} observing windows, there is an indication of stronger optical/IR polarization for \textit{IXPE~2} and \textit{IXPE~3} compared to \textit{IXPE~1}. \textit{IXPE~2} and \textit{3} also exhibit a ratio between the optical/IR and X-ray polarization degree that is significantly higher. In the configuration of an energy stratified jet, it possibly indicates that the optical emission originates from regions that are closer to the shock where the magnetic field is more ordered, i.e. closer to the X-ray emitting region, while for \textit{IXPE~1} the optical flux is emitted further downstream in the jet.\par

By exploiting data from the entire MWL campaign, we find a positive correlation at the level of 4$\,\sigma$ between X-rays and VHE gamma rays without any time delay between both MAGIC energy bands and the 2-10$\,$keV band of \textit{Swift}-XRT. The correlation is at the level of 3$\,\sigma$ with the 0.3-2$\,$keV band. The positive correlation without time-lag supports leptonic scenarios in which the same electron population produces the X-ray and VHE emission, via the synchrotron self-Compton process. Positive correlation at zero time lag were also reported in several previous studies~\citep{2021A&A...655A..89M, 2021A&A...647A..88A, 2021MNRAS.504.1427A,2015A&A...576A.126A}. Such a positive correlation suggests that VHE gamma rays are also emitted close to the shock front (co-spatially to the X-rays). The higher significance obtained when using the X-ray 2-10\,keV band instead of the 0.3-2\,keV band suggests that the VHE emission has a tighter relation with the X-ray fluxes above a few keVs rather than below that. Looking at Fig.~\ref{fig:SED}, this implies that the falling edge of the high-energy SED component is mostly dominated by electrons that emit synchrotron photons well above $\nu_p$, which is in agreement with the expectation of leptonic scenarios \citep{1998ApJ...509..608T}.\par 

At lower energies, we find a marginal evidence of anti-correlation between the X-ray and UV fluxes from May 2022 to June 2022. In this time span, while the X-ray emission shows a long-term flux decay and spectral softening, the UV emission is rising in a quasi monotonic trend. We find that the marginal evidence of correlation happens at zero time lag, without any indication of a delay. Although the significance is estimated $\sim2.5\sigma$ using Monte Carlo simulations, this suggestion is interesting in the context of previous results as well as the newly available X-ray polarization measurements. First, we stress that it is the third time that an indication of X-ray/UV anti-correlation is reported in Mrk~421 \citep{2015A&A...576A.126A,2021A&A...655A..89M}, and each previous indication displays a similar anti-correlation trend over $\sim$monthly timescales. Secondly, the direct implication of an anti-correlation is a physical connection between the X-ray and UV/optical emitting regions. While the \textit{IXPE} results strongly suggest that those regions are not co-spatial, the anti-correlation further supports a scenario in which particles are first accelerated close to a shock front and then advect (and cool) towards a broader region in the jet and dominate the observed UV/optical emission.\par 

A possible scenario explaining the anti-correlation is a long-term evolution of the acceleration efficiency while the electron injection luminosity stays roughly constant. In the latter configuration, a decrease of the acceleration efficiency would increase the relative proportion of lower-energy electrons and shift the synchrotron SED towards lower frequencies (as suggested by the data), while keeping the amplitude of the SED peak at a roughly similar level. This scenario is thus expected to generate an increase of the UV/optical flux (rising edge of the synchrotron component) and a decrease of the X-ray flux (falling edge of the synchrotron component).\par

\citet{2016MNRAS.463.3365A} found an indication of anti-correlation between the optical polarization degree and the synchrotron peak frequencies $\nu_p$ for a sample of BL Lac objects. This behavior was qualitatively explained by the fact that, in the case of BL Lac objects with lower $\nu_p$ (as LSPs), the synchrotron peak is close to the optical band, which is emitted by freshly accelerated electrons near the shock. For HSPs, the optical range is farther from $\nu_p$, and thus comprises emission radiated by electrons that had time to advect away. It is downstream from the shock, where the level of magnetic field disorder increases thus reducing the observed optical polarization degree. In the case where the anti-correlation between the UV and X-rays described above is caused by a shift of $\nu_p$ towards lower frequencies, one would thus expect a simultaneous rise of the optical polarization degree over time, with a value approaching to one in the X-rays. Consistently, the period during which we report an indication of anti-correlation is accompanied by an increase of the optical polarization degree (see Sect.~\ref{sec:optical_pol_evol}). The higher optical polarization degree would also explain the relatively high ratio between the optical/IR and X-ray polarization degree throughout the \textit{IXPE~2} and \textit{IXPE~3} epochs (which are within the time range where a hint of UV/X-ray anti-correlation is reported).\par

Alternatively, the rise of the optical polarization degree during the UV/X-ray anti-correlation time range may be caused by a progressive increase of the relative dominance of a few emitting zones radiating the optical/UV flux. Indeed, in the case where the optical flux receives contributions from many regions with different magnetic field configurations, the polarization degree would decrease.\par

\begin{acknowledgements}

We would like to thank the Instituto de Astrof\'{\i}sica de Canarias for the excellent working conditions at the Observatorio del Roque de los Muchachos in La Palma. The financial support of the German BMBF, MPG and HGF; the Italian INFN and INAF; the Swiss National Fund SNF; the grants PID2019-104114RB-C31, PID2019-104114RB-C32, PID2019-104114RB-C33, PID2019-105510GB-C31, PID2019-107847RB-C41, PID2019-107847RB-C42, PID2019-107847RB-C44, PID2019-107988GB-C22 funded by the Spanish MCIN/AEI/ 10.13039/501100011033; the Indian Department of Atomic Energy; the Japanese ICRR, the University of Tokyo, JSPS, and MEXT; the Bulgarian Ministry of Education and Science, National RI Roadmap Project DO1-400/18.12.2020 and the Academy of Finland grant nr. 320045 is gratefully acknowledged. This work was also been supported by Centros de Excelencia ``Severo Ochoa'' y Unidades ``Mar\'{\i}a de Maeztu'' program of the Spanish MCIN/AEI/ 10.13039/501100011033 (SEV-2016-0588, CEX2019-000920-S, CEX2019-000918-M, CEX2021-001131-S, MDM-2015-0509-18-2) and by the CERCA institution of the Generalitat de Catalunya; by the Croatian Science Foundation (HrZZ) Project IP-2016-06-9782 and the University of Rijeka Project uniri-prirod-18-48; by the Deutsche Forschungsgemeinschaft (SFB1491 and SFB876); the Polish Ministry Of Education and Science grant No. 2021/WK/08; and by the Brazilian MCTIC, CNPq and FAPERJ. A.A.E. and D.P. acknowledge support from the Deutsche Forschungsgemeinschaft (DFG; German Research Foundation) under Germany’s Excellence Strategy EXC-2094—390783311.

The Imaging X-ray Polarimetry Explorer ({\it IXPE}) is a joint US and Italian mission. 
The US contribution is supported by the National Aeronautics and Space Administration (NASA) and led and managed by its Marshall Space Flight Center (MSFC), with industry partner Ball Aerospace (contract NNM15AA18C). The Italian contribution is supported by the Italian Space Agency (Agenzia Spaziale Italiana, ASI) through contract ASI-OHBI-2017-12-I.0, agreements ASI-INAF-2017-12-H0 and ASI-INFN-2017.13-H0, and its Space Science Data Center (SSDC), and by the Istituto Nazionale di Astrofisica (INAF) and the Istituto Nazionale di Fisica Nucleare (INFN) in Italy. This research used data products provided by the {\it IXPE} Team (MSFC, SSDC, INAF, and INFN) and distributed with additional software tools by the High-Energy Astrophysics Science Archive Research Center (HEASARC), at NASA Goddard Space Flight Center (GSFC).

The \textit{Fermi} LAT Collaboration acknowledges generous ongoing support
from a number of agencies and institutes that have supported both the
development and the operation of the LAT as well as scientific data analysis.
These include the National Aeronautics and Space Administration and the
Department of Energy in the United States, the Commissariat \`a l'Energie Atomique
and the Centre National de la Recherche Scientifique / Institut National de Physique
Nucl\'eaire et de Physique des Particules in France, the Agenzia Spaziale Italiana
and the Istituto Nazionale di Fisica Nucleare in Italy, the Ministry of Education,
Culture, Sports, Science and Technology (MEXT), High Energy Accelerator Research
Organization (KEK) and Japan Aerospace Exploration Agency (JAXA) in Japan, and
the K.~A.~Wallenberg Foundation, the Swedish Research Council and the
Swedish National Space Board in Sweden.
 
Additional support for science analysis during the operations phase is gratefully
acknowledged from the Istituto Nazionale di Astrofisica in Italy and the Centre
National d'\'Etudes Spatiales in France. This work performed in part under DOE
Contract DE-AC02-76SF00515.

The IAA-CSIC group acknowledges financial support from the Spanish ``Ministerio de Ciencia e Innovación'' (MCIN/AEI/ 10.13039/501100011033) through the Center of Excellence Severo Ochoa award for the Instituto de Astrof\'{i}sica de Andaluc\'{i}a-CSIC (CEX2021-001131-S), and through grants PID2019-107847RB-C44 and PID2022-139117NB-C44. The POLAMI observations were carried out at the IRAM 30m Telescope. IRAM is supported by INSU/CNRS (France), MPG (Germany), and IGN (Spain). Some of the data are based on observations collected at the Observatorio de Sierra Nevada, owned and operated by the Instituto de Astrof\'{i}sica de Andaluc\'{i}a (IAA-CSIC). Further data are based on observations collected at the Centro Astron\'{o}mico Hispano en Andalucía (CAHA), operated jointly by Junta de Andaluc\'{i}a and Consejo Superior de Investigaciones Cient\'{i}ficas (IAA-CSIC). 

Some of the data reported here are based on observations made with the Nordic Optical Telescope, owned in collaboration with the University of Turku and Aarhus University, and operated jointly by Aarhus University, the University of Turku, and the University of Oslo, representing Denmark, Finland, and Norway, the University of Iceland and Stockholm University at the Observatorio del Roque de los Muchachos, La Palma, Spain, of the Instituto de Astrofisica de Canarias. E. L. was supported by Academy of Finland projects 317636 and 320045.  The data presented here were obtained (in part) with ALFOSC, which is provided by the Instituto de Astrofisica de Andalucia (IAA) under a joint agreement with the University of Copenhagen and NOT. We acknowledge funding to support our NOT observations from the Finnish Centre for Astronomy with ESO (FINCA), University of Turku, Finland (Academy of Finland grant nr 306531). We are grateful to Vittorio Braga, Matteo Monelli, and Manuel S\"{a}nchez Benavente for performing the observations at the Nordic Optical Telescope.

The research at Boston University was supported in part by National Science Foundation grant AST-2108622, NASA Fermi Guest Investigator grants 80NSSC21K1917 and 80NSSC22K1571, and NASA Swift Guest Investigator grant 80NSSC22K0537. This research was conducted in part using the Mimir instrument, jointly developed at Boston University and Lowell Observatory and supported by NASA, NSF, and the W.M. Keck Foundation. We thank D.~Clemens for guidance in the analysis of the Mimir data. 

This research has made use of data from the RoboPUol program, a collaboration between Caltech, the University of Crete, IA-FORTH, IUCAA, the MPIfR, and the Nicolaus Copernicus University, which was conducted at Skinakas Observatory in Crete, Greece. D.B., S.K., R.S., N.M., acknowledge support from the European Research Council (ERC) under the European Unions Horizon 2020 research and innovation program under grant agreement No. 771282. C.C. acknowledges support from the European Research Council (ERC) under the HORIZON ERC Grants 2021 program under grant agreement No. 101040021. 

This study used observations conducted with the 1.8m Perkins Telescope (PTO) in Arizona (USA), which is owned and operated by Boston University. This work was supported by NSF grant AST-2109127. 

We acknowledge the use of public data from the Swift data archive. Based on observations obtained with XMM-Newton, an ESA science mission with instruments and contributions directly funded by ESA Member States and NASA.

This publication makes use of data obtained at Mets\"ahovi Radio Observatory, operated by Aalto University in Finland. The Submillimetre Array is a joint project between the Smithsonian Astrophysical Observatory and the Academia Sinica Institute of Astronomy and Astrophysics and is funded by the Smithsonian Institution and the Academia Sinica. Mauna Kea, the location of the SMA, is a culturally important site for the indigenous Hawaiian people; we are privileged to study the cosmos from its summit. This work was supported by JST, the establishment of university fellowships towards the creation of science and technology innovation, Grant Number JPMJFS2129. This work was supported by Japan Society for the Promotion of Science (JSPS) KAKENHI Grant Numbers JP21H01137. This work was also partially supported by the Optical and Near-Infrared Astronomy Inter-University Cooperation Program from the Ministry of Education, Culture, Sports, Science and Technology (MEXT) of Japan. We are grateful to the observation and operating members of the Kanata Telescope.  

S.K. acknowledges support from the European Research Council (ERC) under the European Unions Horizon 2020 research and innovation program under grant agreement No.~771282.

\end{acknowledgements}

\bibliographystyle{aa}
\bibliography{bibliography_paper}

\begin{appendix}

\section{\textit{XMM-Newton} fine-binned light curves}
\label{sec:appendix_xmm_light_curve}

\begin{figure*}[h!]
\centering
  \resizebox{\hsize}{!}{\includegraphics{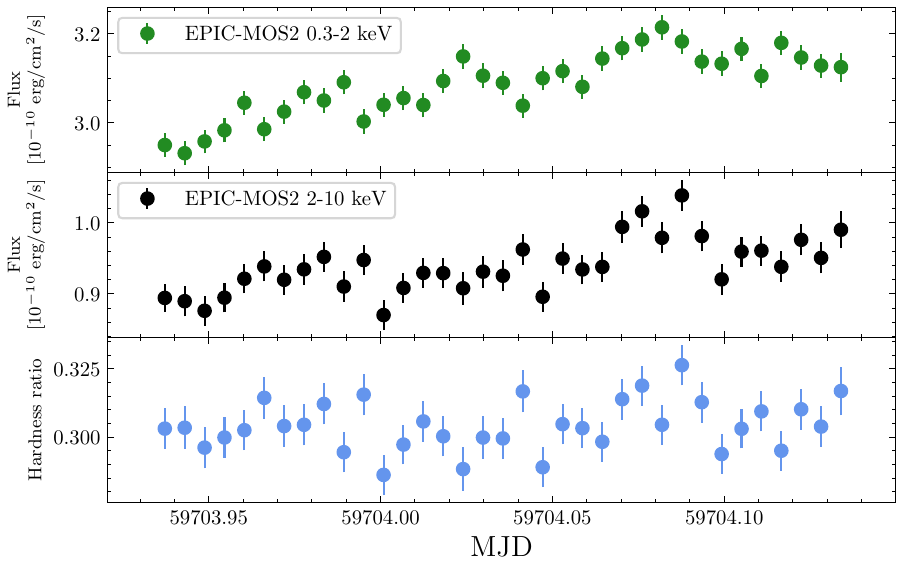}}
  \caption{\textit{XMM-Newton} light curve from the EPIC-MOS2 camera in the 0.3-2\,keV and 2-10\,keV bands during the \textit{IXPE~1} epoch. The lower panel is the ratio between the 2-10\,keV and 0.3-2\,keV fluxes.}
  \label{fig:xmm_lc_ixpe1}
\end{figure*}

\begin{figure*}[h!]
\centering
  \resizebox{\hsize}{!}{\includegraphics{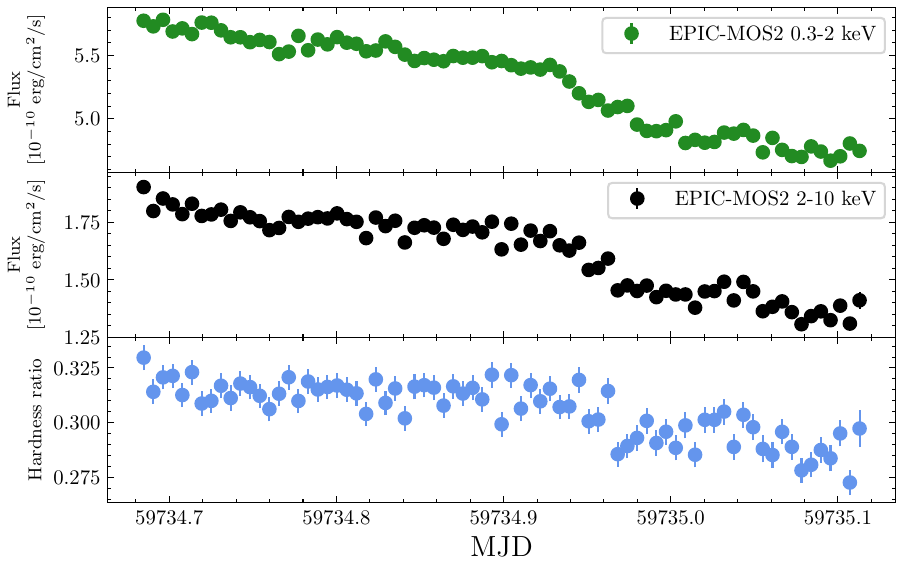}}
   \caption{\textit{XMM-Newton} light curve from the EPIC-MOS2 camera in the 0.3-2\,keV and 2-10\,keV bands during the \textit{IXPE~2} epoch. The lower panel is the ratio between the 2-10\,keV and 0.3-2\,keV fluxes.}
  \label{fig:xmm_lc_ixpe2}
\end{figure*}

In this section, we present the fine-binned light curves from both of the \textit{XMM-Newton} exposures analyzed in this work. The fluxes are computed using the EPIC-MOS2 camera (which has the largest exposure time among the operating instruments on board \textit{XMM-Newton}) in the 0.3-2\,keV and 2-10\,keV bands using a temporal binning of 500\,s.  The SAS task {\tt epiclccorr} is used to produce background subtracted source count rates corrected for inefficiencies of the instrument (vignetting, chip gaps, PSF...) and time corrThe modelling yieldsections (dead time, GTIs...). The count rates are then converted to energy fluxes (i.e. in erg/cm$^{2}$/s units) assuming the best-fit log parabola model over the entire exposure. The light curve for the \textit{IXPE~1} and the \textit{IXPE~2} epochs are shown in Fig.~\ref{fig:xmm_lc_ixpe1} and Fig.~\ref{fig:xmm_lc_ixpe2}, respectively. The bottom panel present the ratio between the 2-10\,keV and the 0.3-2\,keV fluxes.


\clearpage

\section{VHE versus 0.2-3\,keV DCF analysis}
\label{sec:appendix_vhe_vs_xray}
\begin{figure}[h!]
\centering
  \resizebox{\hsize}{!}{\includegraphics{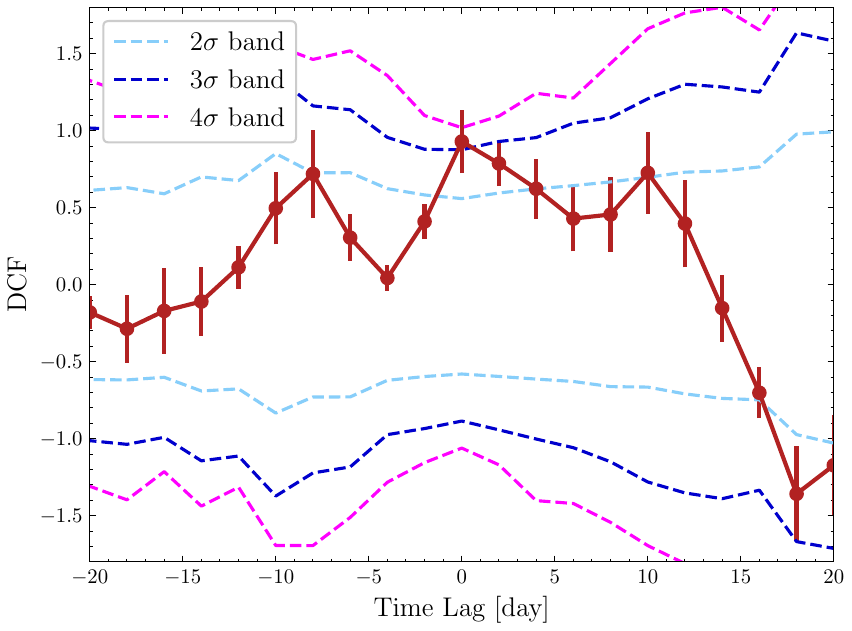}}
  \caption{Discrete correlation function DCF computed for the MAGIC $0.2-1$\,TeV and \textit{Swift}-XRT 0.3-2$\,$keV light curves between MJD~59700 (May 1\textsuperscript{st} 2022) and MJD~59740 (June 10\textsuperscript{th} 2022) with a time binning of 2 days. The red points are the obtained DCF values and their uncertainties. The light blue, dark blue and pink dashed lines show the $2\sigma$, $3\sigma$ and $4\sigma$ significance bands, respectively (see text for more details).}
  \label{fig:DCF_magic_LE_Xray_le}
\end{figure}

\begin{figure}[h!]
\centering
  \resizebox{\hsize}{!}{\includegraphics{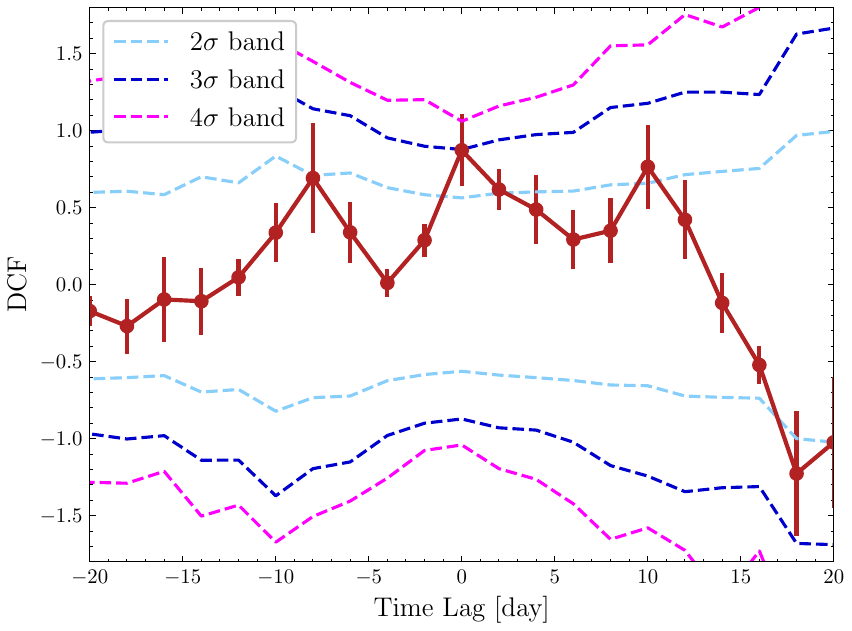}}
  \caption{Discrete correlation function DCF computed for the MAGIC $>1$\,TeV and \textit{Swift}-XRT 0.3-2$\,$keV light curves between MJD~59700 (May 1\textsuperscript{st} 2022) and MJD~59740 (June 10\textsuperscript{th} 2022) with a time binning of 2 days. The red points are the obtained DCF values and their uncertainties. The light blue, dark blue and pink dashed lines show the $2\sigma$, $3\sigma$ and $4\sigma$ significance bands, respectively (see text for more details).}
  \label{fig:DCF_magic_HE_Xray_le}
\end{figure}

This section presents the results of the DCF analysis between the VHE fluxes and the 0.3-2\,keV band from \textit{Swift}-XRT. Fig.~\ref{fig:DCF_magic_LE_Xray_le} and Fig.~\ref{fig:DCF_magic_HE_Xray_le} show the DCF when the VHE flux is computed in the 0.2-1\,TeV and $>1$\,TeV ranges, respectively. The dashed lines are the confidence bands based on Monte Carlo simulations (see Sect.~\ref{sec:long_term_MWL} for more details).

\clearpage

\section{UV versus X-ray correlation}
\label{sec:appendix_uv_vs_xray}
\begin{figure}[h!]
\centering
  \resizebox{\hsize}{!}{\includegraphics{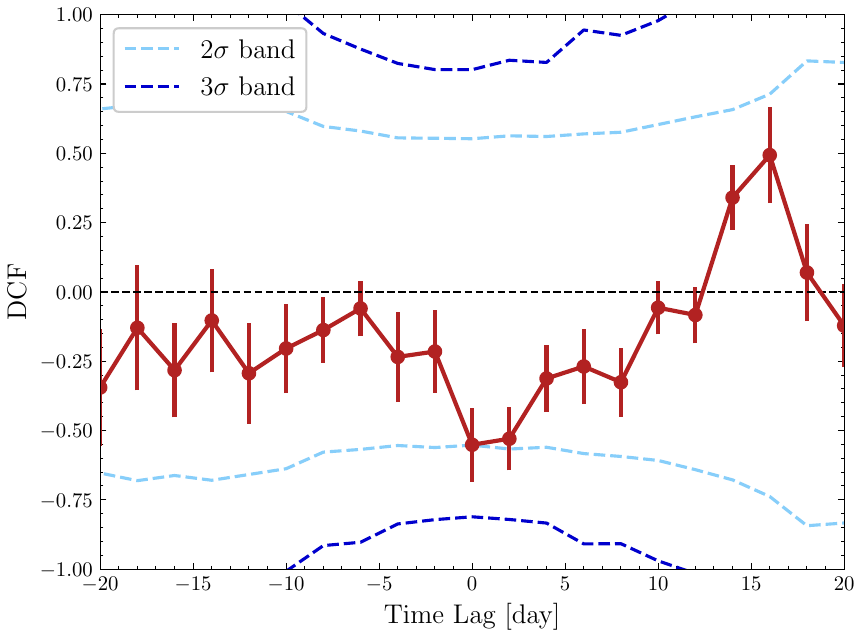}}
  \caption{Discrete correlation function DCF computed for the \textit{Swift}-UVOT W1 and \textit{Swift}-XRT 0.3-2$\,$keV light curves over the second part of the MWL campaign, between MJD~59710 (May 11\textsuperscript{th} 2022) and MJD~59760 (June 30\textsuperscript{th} 2022), with a time-lag binning of 2\,days. The red points are the obtained DCF values and their uncertainties. The light blue and dark blue dashed lines show the $2\sigma$ and $3\sigma$ significance bands, respectively (see text for more details).}
  \label{fig:DCF_UV_Xray_le}
\end{figure}

\begin{figure}[h!]
\centering
  \resizebox{\hsize}{!}{\includegraphics{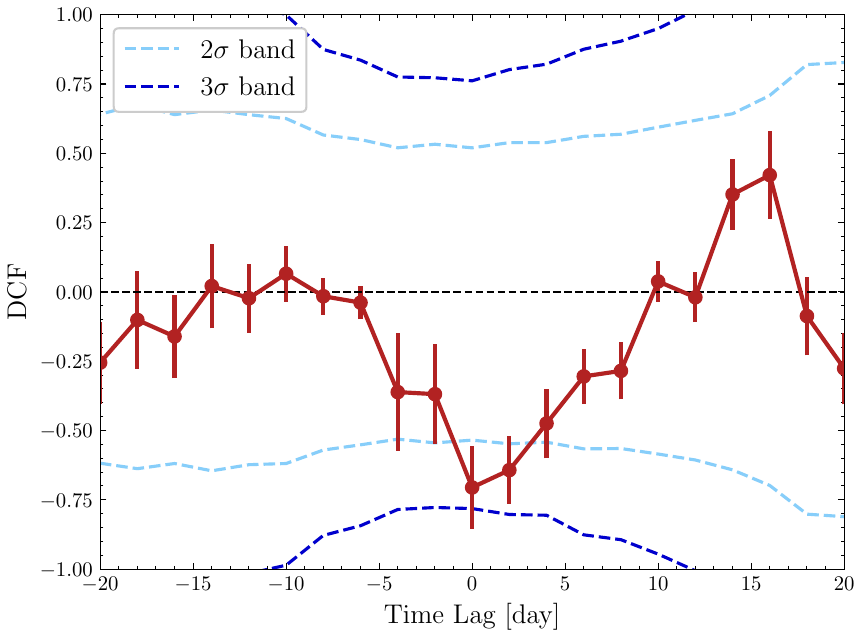}}
  \caption{Discrete correlation function DCF computed for the \textit{Swift}-UVOT W1 and \textit{Swift}-XRT 2-10$\,$keV light curves over the second part of the MWL campaign, between MJD~59710 (May 11\textsuperscript{th} 2022) and MJD~59760 (June 30\textsuperscript{th} 2022), with a time-lag binning of 2\,days. The red points are the obtained DCF values and their uncertainties. The light blue and dark blue dashed lines show the $2\sigma$ and $3\sigma$ significance bands, respectively (see text for more details).}
  \label{fig:DCF_UV_Xray_he}
\end{figure}

\begin{figure}[h!]
\centering
  \resizebox{\hsize}{!}{\includegraphics{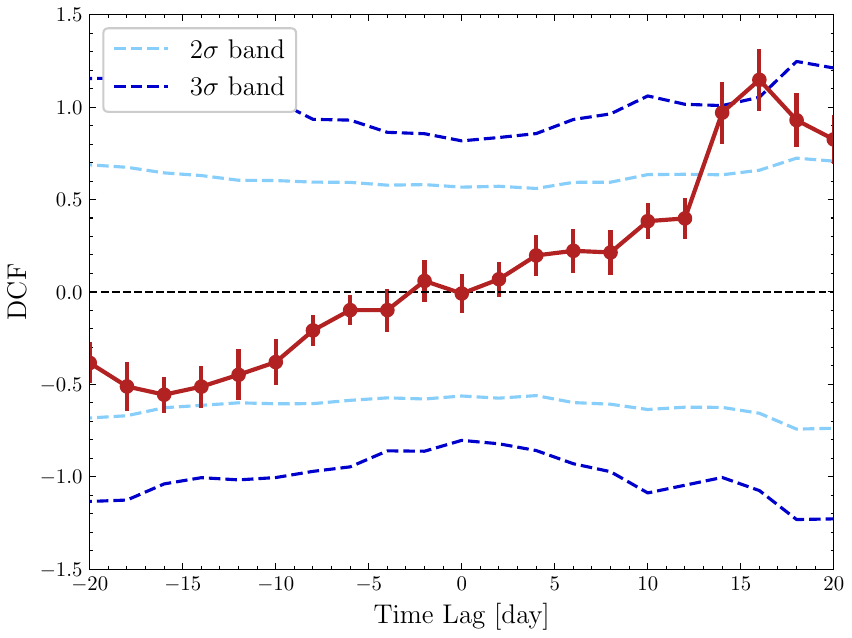}}
  \caption{Discrete correlation function DCF computed for the \textit{Swift}-UVOT W1 and \textit{Swift}-XRT 0.3-2$\,$keV light curves over the full MWL campaign, between MJD~59695 (April 26\textsuperscript{th} 2022) and MJD~59760 (June 30\textsuperscript{th} 2022), with a time-lag binning of 2\,days. The red points are the obtained DCF values and their uncertainties. The light blue and dark blue dashed lines show the $2\sigma$ and $3\sigma$ significance bands, respectively (see text for more details).}
  \label{fig:DCF_UV_Xray_le_full_camp}
\end{figure}

\begin{figure}[h!]
\centering
  \resizebox{\hsize}{!}{\includegraphics{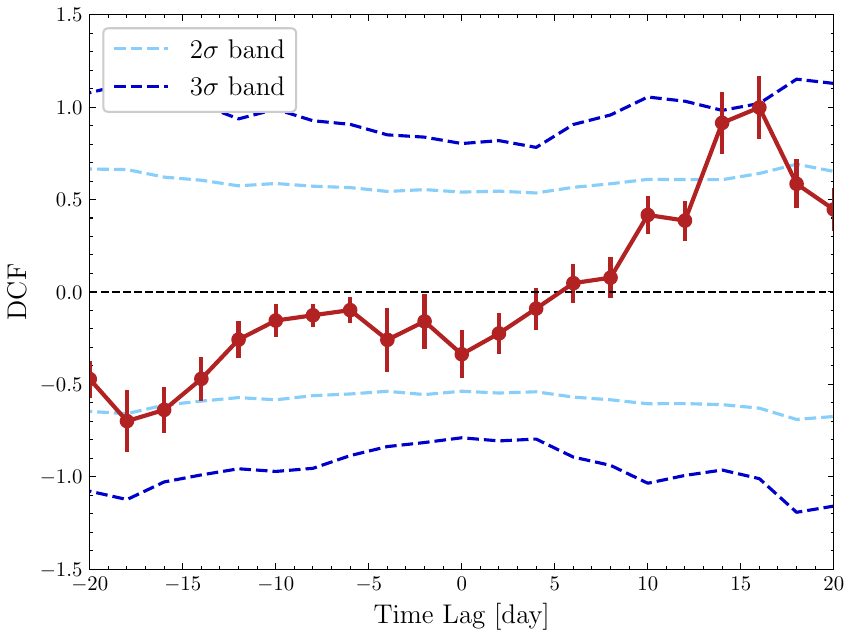}}
  \caption{Discrete correlation function DCF computed for the \textit{Swift}-UVOT W1 and \textit{Swift}-XRT 2-10$\,$keV light curves over the full campaign, between MJD~59695 (April 26\textsuperscript{th} 2022) to MJD~59760 (June 30\textsuperscript{th} 2022), with a time-lag binning of 2\,days. The red points are the obtained DCF values and their uncertainties. The light blue and dark blue dashed lines show the $2\sigma$ and $3\sigma$ significance bands, respectively (see text for more details).}
  \label{fig:DCF_UV_Xray_he_full_camp}
\end{figure}

This section presents the results of the DCF analysis between the X-ray and the UV fluxes in the \textit{Swift}-UVOT W1 filter. Fig.~\ref{fig:DCF_UV_Xray_le} and Fig.~\ref{fig:DCF_UV_Xray_he} show the DCF when the X-ray flux is computed in the 0.3-2\,keV and 2-10\,keV ranges, respectively, using data between MJD~59710 and MJD~59740 (i.e., corresponding to the second part of the MWL campaign presented in this work; May 11\textsuperscript{th} 2022 to June 10\textsuperscript{th} 2022). The dashed lines are the confidence bands based on Monte Carlo simulations (see Sect.~\ref{sec:long_term_MWL} for more details about the procedure). In Fig.~\ref{fig:DCF_UV_Xray_le_full_camp} and Fig.~\ref{fig:DCF_UV_Xray_he_full_camp}, we display the results after repeating the exercise when data from the entire MWL campaign were included (i.e., from MJD~59695 to MJD~59740; April 26\textsuperscript{th} 2022 to June 10\textsuperscript{th} 2022).



\clearpage

\section{Modelling of the X-ray and VHE spectra during the \textit{IXPE} epochs.}
\label{sec:modelling_xray_vhe}

In Sect.~\ref{sec:discussion}, we present calculations of the ratio between the cooling and acceleration timescales of X-ray-emitting electrons during the hysteresis loops that we detect in the \textit{NuSTAR} data during \textit{IXPE~3}.  The ratio between the two timescales is then used to estimate the \textit{gyro-factor} $\xi$. For simplification, only synchrotron cooling is considered, and IC cooling is neglected. In this section, we address the validity of this assumption.\par 

Following the notation of \citet{1998ApJ...509..608T}, the IC cooling timescale is estimated as :
\begin{equation}
\label{eq:ic_cooling_formula}
t'_{cool, IC} = \frac{3 m_e c}{4 \sigma_T U'_{synch, avail} \gamma'} 
\end{equation}
where $U'_{synch, avail}$ is the available target photon density for IC process (below the Klein-Nishina limit - see Eq.~20 in \citet{1998ApJ...509..608T}) within the emitting zone. The estimation of $t'_{cool, IC}$ requires the knowledge of $U'_{synch, avail}$, which we extract with a simple modeling of the SED by considering a one-zone SSC model \citep{1992ApJ...397L...5M, 1999ApJ...521..145M}. For this exercise, we aim at describing the X-ray and VHE spectra only for the following reasons. First, a description of the radio-to-VHE data would require a more complex modelling that takes into account the energy-stratification of the jet implied by the broadband polarization data. This effort lies beyond the scope of this work. Secondly, describing the X-ray \& VHE spectra in a one-zone SSC approach is motivated by the tight X-ray/VHE correlation at zero time-lag. Since only \textit{IXPE~1} and \textit{IXPE~2} have simultaneous X-ray \& VHE data we are forced to focus on those two epochs to constrain physical parameters of the source during \textit{IXPE~3}, where the hysteresis loops actually happened. This represents a caveat for the following analysis since the source parameters may have evolved between the different epochs.\par

\begin{figure}[h!]
        \centering
        \begin{subfigure}[b]{0.95\columnwidth}
            \centering
            \includegraphics[width=1.03\linewidth]{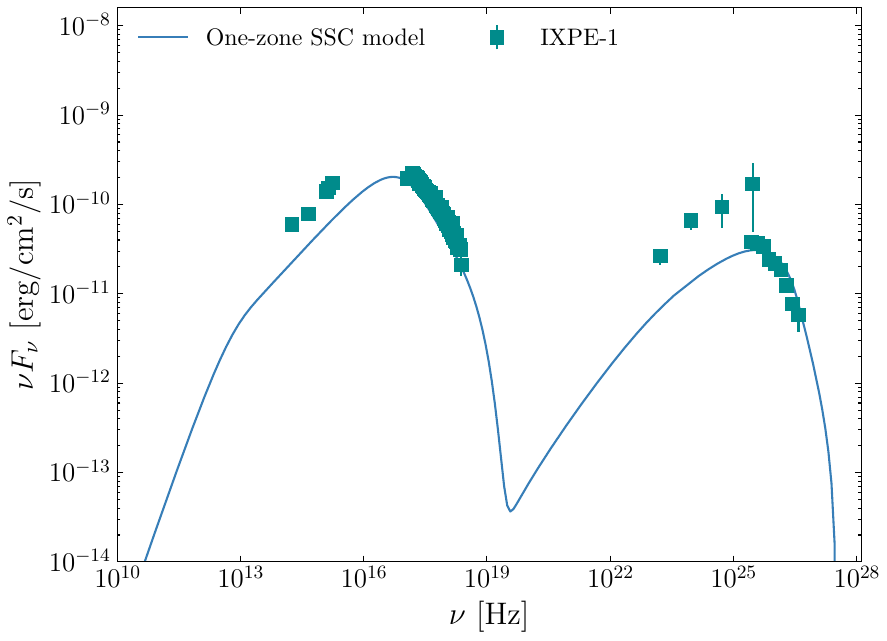}
        \end{subfigure}
        \begin{subfigure}[b]{0.95\columnwidth}  
            \centering 
            \includegraphics[width=1.03\linewidth]{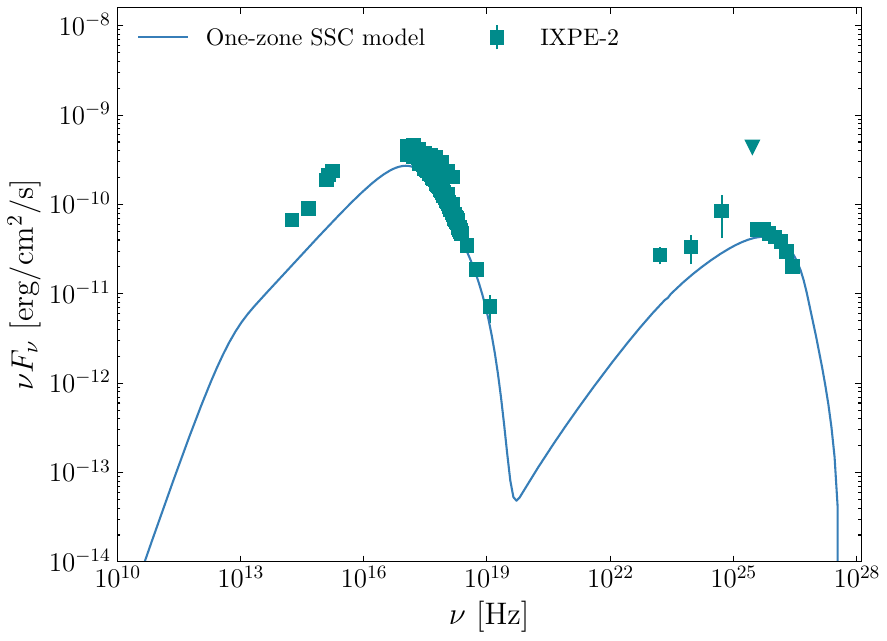}
        \end{subfigure}
        \caption{Results of a one-zone SSC model applied to the \textit{IXPE~1} (top figure) and \textit{IXPE~2} (bottom figure) epochs in order to constrain the physical parameters of the X-ray \& VHE emitting region. The data are plotted with cyan markers, and the model is shown as a solid blue line. The obtained modelling parameters are listed in Table~\ref{tab:SED_model_IXPE1} and Table~\ref{tab:SED_model_IXPE2}. The reader is referred to Sect.~\ref{sec:modelling_xray_vhe} for more details on the model.}  

\label{fig:IXPE_onezones}
\end{figure}

We first fix the radius of the emitting region to $R' = 2 \times 10^{16}$\,cm. It is derived from the constraints using causality arguments, $R' \lesssim \delta \cdot c \cdot t_{var,obs}$ \citep{1998ApJ...509..608T}, where $t_{var,obs}$ is the observed variability timescale and $\delta$ the Doppler factor that we fix to 30 \citep[which is a typical value adopted for Mrk~421 in previous modelling, see e.g.][]{1998ApJ...509..608T, 2016ApJ...819..156B, 2021A&A...655A..89M}. Here, we set $t_{var,obs}=7$\,hr, which is the halving/doubling time that we measure in the \textit{NuSTAR} band. We model the electron distribution with a broken power-law,
\begin{equation}
    \frac{dN'}{d\gamma'}(\gamma')= \begin{cases}
    N'_0\, \gamma'{}^{-n_1}, \quad \gamma'_{min}<\gamma'<\gamma'_{br}\\
    N'_0\, \gamma_{br}'{}^{n_2-n_1} \gamma'^{-n_2}, \quad \gamma'_{br}<\gamma'<\gamma'_{max}\, ,\\
    \end{cases}
\end{equation}
\noindent
where $N'_0$ is a normalisation constant. $\gamma'_{min}$, $\gamma'_{br}$, and $\gamma'_{max}$ are defined as the minimum, break, and maximum Lorentz factor, respectively. Differently from $n_2$, $n_1$ can not be constrained by the X-ray \& VHE data, so we fix $n_1=2.0$, close to the predictions of shock acceleration \citep{2000ApJ...542..235K}. The overall electron energy density is given by $U'_{e}$. The resulting models are shown in Fig.~\ref{fig:IXPE_onezones}, and exhibit a reasonable description of the X-ray \& VHE data. We list in Table~\ref{tab:SED_model_IXPE1} and Table~\ref{tab:SED_model_IXPE2} the obtained parameters. The optical/UV and MeV-GeV data are purposely underpredicted. In fact, the energy-stratification of jet suggested by the polarization data strongly implies that optical/UV and MeV-GeV fluxes receive a significant contribution from broader and separate regions than the X-ray and VHE one. Hence, our one-zone modelling does not intend to describe the entire SED.\par 

The modelling yields $U'_{synch, avail} < U'_{B}$ in both epochs. From Eq.~\ref{eq:ic_cooling_formula} and Eq.~\ref{eq:synch_cooling_formula}, one thus concludes that IC cooling timescale is longer than the synchrotron cooling timescale. Only considering synchrotron cooling is thus a reasonable simplification to assess the cooling dynamics of the electrons during the hysteresis loops that we report and discuss in Sect.~\ref{sec:intranight_full_nustar} \& Sect.~\ref{sec:discussion}.

\begin{table}[h!]
\caption{\label{tab:SED_model_IXPE1} Model parameters of the one-zone SSC model applied to the \textit{IXPE 1} epoch.} 
\centering
\begin{tabular}{ l c c }     
\hline\hline 
 Parameter & Value \\  
\hline\hline   
$B'$ [G] & $4.2 \times 10^{-2}$ \\
$R'$ [cm] & $2\times10^{16}$ \\
$\delta$ & 30 \\
$U'_{e}$ [erg\,cm$^{-3}$] & $9.5\times10^{-4}$ \\
$n_1$ & 2.0 \\
$n_2$ & 4.5 \\
$\gamma'_{min}$ & $10^{3}$ \\
$\gamma'_{br}$ & $1.1 \times 10^{5}$ \\
$\gamma'_{max}$ & $0.9 \times 10^{6}$ \\
\hline
$U'_{B}$ [erg\,cm$^{-3}$] & $0.7\times10^{-4}$ \\
$U'_{synch, avail}$ [erg\,cm$^{-3}$] & $0.3\times10^{-4}$ \\
\hline
\end{tabular}
\tablefoot{\centering See text in Appendix~\ref{sec:modelling_xray_vhe} for a description of the parameters.} 
\end{table}

\begin{table}[h!]
\caption{\label{tab:SED_model_IXPE2} Model parameters of the one-zone SSC model applied to the \textit{IXPE 2} epoch.} 
\centering
\begin{tabular}{ l c c }     
\hline\hline 
 Parameter & Value \\  
\hline\hline   
$B'$ [G] & $3.8 \times 10^{-2}$ \\
$R'$ [cm] & $2\times10^{16}$ \\
$\delta$ & 30 \\
$U'_{e}$ [erg\,cm$^{-3}$] & $11.0\times10^{-4}$ \\
$n_1$ & 2.0 \\
$n_2$ & 4.7 \\
$\gamma'_{min}$ & $10^{3}$ \\
$\gamma'_{br}$ & $1.8 \times 10^{5}$ \\
$\gamma'_{max}$ & $1.1 \times 10^{6}$ \\
\hline
$U'_{B}$ [erg\,cm$^{-3}$] & $0.6\times10^{-4}$ \\
$U'_{synch, avail}$ [erg\,cm$^{-3}$] & $0.3\times10^{-4}$ \\
\hline
\end{tabular}
\tablefoot{\centering See text in Appendix~\ref{sec:modelling_xray_vhe} for a description of the parameters.} 
\end{table}

\end{appendix}

\end{document}